\documentclass[
  aps,
  prx,
  reprint,
  superscriptaddress,
  amsmath,
  amssymb,
  longbibliography,
  nofootinbib,
  floatfix
]{revtex4-2}

\usepackage{graphicx}
\usepackage{dcolumn}
\usepackage{bm}
\usepackage{booktabs}
\usepackage{xurl}
\usepackage{hyperref}
\hypersetup{hidelinks}

\graphicspath{{figures/}}

\newcommand{\FigWorkflowComparisonWidth}{0.98\textwidth}
\newcommand{\FigArchitectureWidth}{0.98\textwidth}
\newcommand{\FigExamplePromptWidth}{0.98\textwidth}
\newcommand{\FigBenchmarkWidth}{\textwidth}
\newcommand{\FigAblationWidth}{\textwidth}
\newcommand{\FigPhotocatalystPdosTopWidth}{0.82\textwidth}
\newcommand{\FigPhotocatalystPdosBottomWidth}{0.7\textwidth}
\newcommand{\FigPhotocatalystDepthWidth}{0.98\columnwidth}
\newcommand{\FigPhotocatalystPldosWidth}{0.98\textwidth}
\newcommand{\FigPldosMaxHeight}{0.58\textheight}
\newcommand{\FigPhotocatalystTrendsWidth}{0.70\textwidth}

\newcommand{\RhInGaN}{Rh$_6$/In$_x$Ga$_{1-x}$N}
\newcommand{\InGaN}{In$_x$Ga$_{1-x}$N}
\newcommand{\Egref}{E_g^{\mathrm{ref}}}
\newcommand{\DEC}{\Delta E_C^{A}}
\newcommand{\DNgap}{\Delta N_{\mathrm{gap}}^{\mathrm{surf}}}
\newcommand{\DNfermi}{\Delta N_{\pm0.1}}

\begin{document}

\title{LatticeMind: A Validation-Gated Agentic Framework for Density Functional Theory Workflows}

\newcommand{\McGillAffil}{Department of Physics, McGill University, Montr\'eal, QC H3A 2T8, Canada}
\newcommand{\NanoAffil}{Nanoacademic Technologies Inc., Montr\'eal, QC H3A 1E7, Canada}

\author{Nathaniel M.\ Vegh}
\email{nathaniel.vegh@mail.mcgill.ca}
\affiliation{\NanoAffil}
\affiliation{\McGillAffil}

\author{Yu Zhu}
\affiliation{\NanoAffil}

\author{Rapha\"el J.\ Prentki}
\affiliation{\NanoAffil}

\author{F\'elix Beaudoin}
\affiliation{\NanoAffil}

\author{Shuaishuai Yuan}
\affiliation{\McGillAffil}

\author{Hong Guo}
\affiliation{\McGillAffil}

\date{July 14, 2026}

\begin{abstract}
First-principles electronic-structure calculations use quantum mechanics to predict material properties and guide materials design, but their reliability depends on the workflow that produces them. A production calculation is rarely a single-stage solver call: it is a multistage workflow in which each stage hands off intermediate results to the next. Language models handle a single-stage calculation quite well but grow unreliable as stages are chained, where a dropped dependency, a wrong file handoff, or a hallucinated input can invalidate the result even when the solver runs successfully. Discovering such errors only after execution wastes expensive computing time and reduces the productivity of materials research. We introduce LatticeMind, a validation-gated agentic framework for density functional theory (DFT) workflows. LatticeMind turns each natural-language request into a fully executable, scientifically checkable DFT project: a workflow contract defines the required stages, files, and dependencies, then deterministic validators test the generated project before DFT execution. Failed checks trigger targeted repair, and final reports remain traceable to the intermediate project files. LatticeMind is benchmarked with two DFT packages: RESCU (Real-space Electronic Structure CalcUlator) and VASP (Vienna \textit{Ab initio} Simulation Package). On a 44-case RESCU benchmark spanning total-energy, density-of-states, band-structure, phonon, relaxation, and band-edge calculations, GPT-5.4-mini, Gemini 3 Flash Preview, and Claude Sonnet 4.6 each reproduced all 44 benchmark reference values under a single scoring standard. Across the 132 total runs, 119 passed on the first validated plan and the remaining 13 passed after one validation-gated repair cycle. An ablation study identified the workflow contract and deterministic validation as the most crucial reliability mechanisms. A conventional trial-and-repair model, representative of the current standard pattern of many agentic systems in which errors are fixed only after failed solver execution, reached 71\% under the same test, versus 100\% for the full LatticeMind agent. On the VASP benchmark, LatticeMind reproduced all reference values in the 19-case subset, showing that the agent architecture is not tied to one solver. As a realistic application, LatticeMind generated 23 RESCU input files for a multistage study of how Rh-induced local electronic responses evolve with In concentration at a \RhInGaN\ photocatalyst interface. This protocol lies outside the benchmark task families, demonstrating that validated workflow construction generalizes to previously unseen research workflows. Together, these results show a practical pattern for auditable AI-assisted scientific computing: workflow contracts, deterministic validation, bounded repair, and traceable project records can substantially improve the reliability of agent-generated first-principles workflows while catching preventable workflow errors before solver execution.
\end{abstract}

\maketitle

\begin{figure*}[t]
\centering
\makebox[\textwidth][c]{\includegraphics[width=\FigWorkflowComparisonWidth]{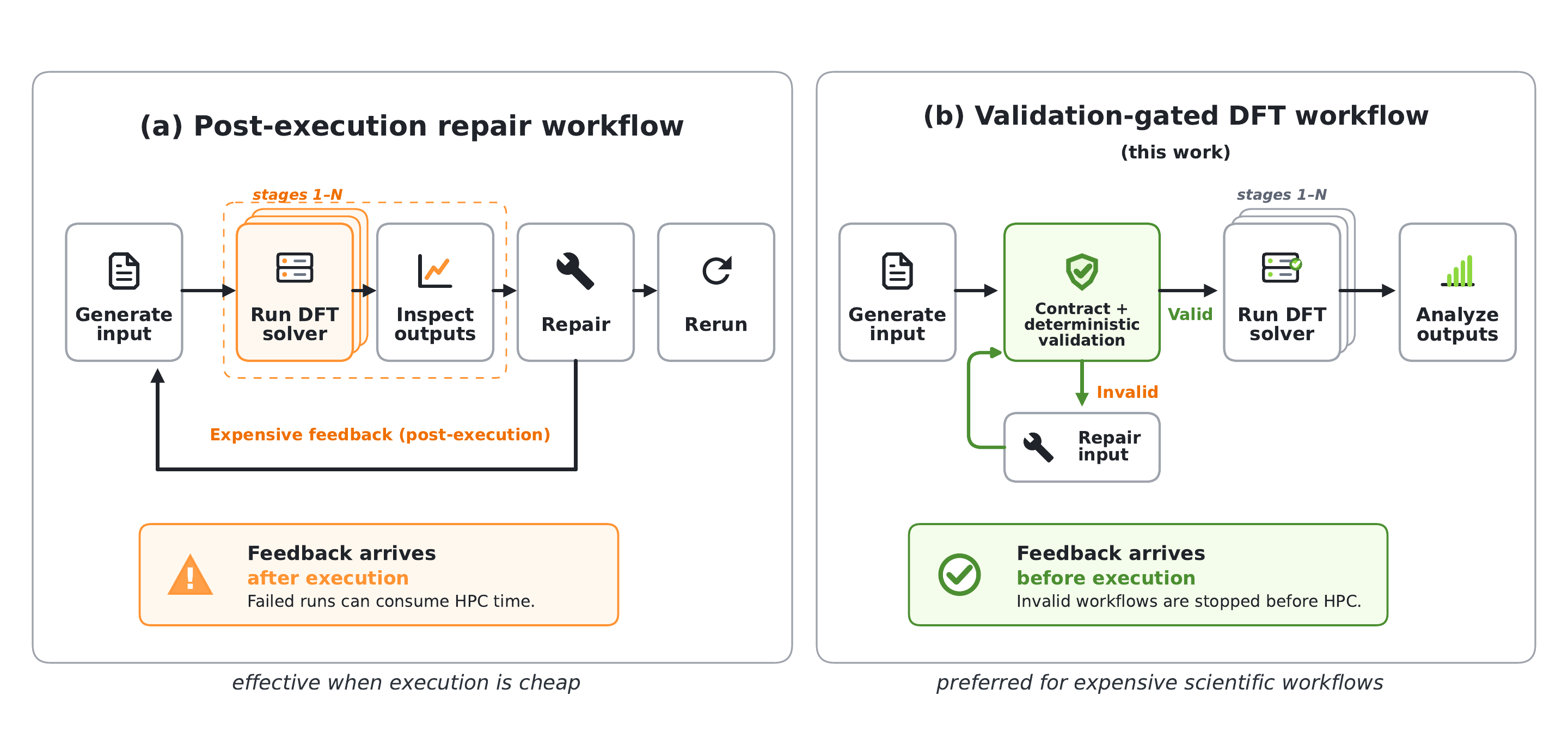}}
\caption{\label{fig:workflow_comparison}
Conceptual comparison between post-execution repair and validation-gated DFT workflow control. (a) In a post-execution repair loop, the agent generates an input, launches the DFT solver, inspects the output or failure, repairs the input, and reruns the calculation. This pattern automates a familiar human trial-and-error workflow, but feedback arrives only after computational resources have been consumed. (b) In LatticeMind, generated inputs and file dependencies are checked against deterministic solver rules and an explicit workflow contract before execution. Invalid projects are repaired or stopped before solver launch, whereas valid projects proceed to approved DFT execution and analysis. The key distinction is that in LatticeMind the DFT solver is no longer the first validator for preventable schema and dependency errors.}
\end{figure*}

\section{Introduction}
First-principles electronic-structure theory, particularly density functional theory (DFT), forms the computational foundation of modern materials design~\cite{Hohenberg1964,Kohn1965,Jones2015DFTReview}. DFT enables predictive calculations of the ground-state properties of materials, including their structural, electronic, magnetic, and vibrational characteristics, while also providing a foundation for evaluating optical responses. Time-dependent and nonequilibrium methods extend DFT to excited states. These capabilities have transformed materials discovery from a largely empirical process into a predictive science, allowing researchers to identify promising materials and optimize their properties prior to synthesis. Due to this usefulness, DFT packages dominate the usage of supercomputers worldwide for materials science calculations. For example, of the 57 million compute hours consumed at the National Energy Research Scientific Computing Center (NERSC) in the United States in 2023~\cite{NERSC2023AnnualReport}, about 20\% were due to a single DFT package~\cite{Zhao2023VASPPerlmutter}. Public materials databases and datasets are built from huge DFT campaigns: the Open Quantum Materials Database (OQMD) reported about 300,000 DFT calculations in 2015~\cite{Kirklin2015OQMD}; a 175k-material dataset required around 10 million core hours~\cite{Schmidt2022PBESolSCAN}; and about 1.4 million DFT calculation tasks were performed to build the MPtrj dataset of more than 1.5 million structures~\cite{Deng2023CHGNet}.  

A DFT-calculated result is typically the output of a multistage workflow: structure preparation, pseudopotential or basis selection, boundary and $k$-point conventions, solver settings, links between stages, and post-processing must all be correct~\cite{Jones2015DFTReview}. An error in any of the stages can invalidate the final answer. In practice, multistage DFT workflows are easy to break even for experienced researchers, so post-execution repair is commonly needed to reach the correct answer, as indicated by the feedback repair loop (black arrow) in Fig.~\ref{fig:workflow_comparison}(a). The repair loop traverses the whole DFT run (orange box) and, for large systems or complicated materials, significant computing resources are wasted. 
Large language models (LLMs) are increasingly used in scientific automation \cite{Boiko2023,Bran2024,Ansari2024,Mirza2025,Mandal2025,Zou2025} to translate scientific intent into workflow steps and help researchers avoid errors. 
For multistage DFT workflows, however, a plausible LLM workflow description is not enough. The scientific object that must be produced is a runnable, validated project: correct input files, properly linked calculation stages, execution logs, validation records, and final outputs that satisfy the predefined workflow checks and yield the correct scientific result. 

Many failures in a DFT run occur not in DFT itself but in the supporting infrastructure around the solver. A stale geometry file from a previous relaxation, an inconsistent pseudopotential path, a missing link to a saved density, or an untracked post-processing convention can invalidate an otherwise reasonable calculation~\cite{Jain2015,Mathew2017,Pizzi2016}. The cause is procedural but the consequence is scientific. At high-throughput scale, it is also a resource problem: a bad file handoff or unsupported input option turns a preventable setup error into a queued solver launch, a failed job record, and another round of expert debugging (Fig.~\ref{fig:workflow_comparison}(a)). 

Reported failure rates of DFT vary by code, workflow, success criterion, and research problem, but published electronic-structure campaigns show that nonproductive execution is common enough to require automation. A molecular DFT workflow for reactive fragments reduced failures from 25.1\% to 1.2\% after automated error correction across more than 12,000 calculations~\cite{Blau2020DFTErrorCorrection}. In the MC3D campaign (the Materials Cloud computational database of inorganic crystals), 67.4\% of workflows completed on first execution, whereas 14.5\% failed despite automated error handling~\cite{Huber2026MC3D}. In a challenging \emph{de novo} transition-metal-complex screen, only 22\% of 6336 DFT geometry optimizations completed successfully under the study's structural-validity criteria~\cite{Gugler2020DeNovoInorganic}. These examples show that preventing avoidable workflow failures is a practical scaling requirement. This is consistent with broader high-performance-computing (HPC) evidence: an empirical study of ten workload traces from eight large-scale HPC systems found that a failed job often consumes more computational resources than a successful one, so failure rates alone understate the machine time lost to failed execution~\cite{Yuan2012HPCFailures}.

This makes first-principles simulation a stringent test for scientific agents. By an agent we mean an LLM-driven system that maintains workflow state, uses tools, executes approved actions, and revises the procedure when checks fail. Recent systems demonstrate the feasibility of this approach in chemistry and materials science, including autonomous experimentation \cite{Boiko2023}, chemistry-tool reasoning \cite{Bran2024}, literature-based knowledge extraction \cite{Ansari2024}, and quantum-chemistry and instrument-control automation \cite{Mandal2025,Zou2025}. Closest in domain to the present work, TritonDFT studies multi-agent automation for Quantum ESPRESSO workflows \cite{Hu2026TritonDFT}. It uses an expert-informed Plan-Execute-Refine workflow and emphasizes Pareto-aware parameter inference, in which the agent selects and refines DFT settings under competing accuracy and computational-cost objectives. GENIUS further demonstrates the value of agentic protocol generation and automated repair for Quantum ESPRESSO, reporting an overall success rate near 80\%, but only 14.2\% zero-shot success and 76.3\% recovery among initially failed cases through an automated error-handling loop~\cite{Soleymanibrojeni2026GENIUS}. These systems show that agentic repair is useful, but they also illustrate the familiar trial-and-repair pattern in which execution feedback remains a central validator, as shown in Fig.~\ref{fig:workflow_comparison}(a). 

\begin{figure}[t]
\centering
\includegraphics[width=\columnwidth]{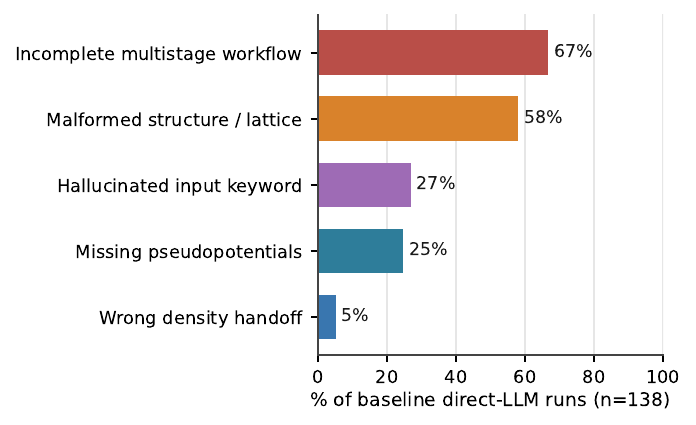}
\caption{\label{fig:llm_failure_modes}
How a bare direct-LLM baseline fails. Bars show the fraction of direct-LLM and LLM-plus-documentation runs on the benchmark below that exhibit each failure mode; a single run can show several. The failures are dominated by incomplete multistage workflows (missing or unlinked downstream stages), malformed structures, hallucinated input fields, missing pseudopotentials, and broken saved-density handoffs, the errors that arise specifically when independent calculations must be chained.}
\end{figure}

The difficulty agentic systems confront is sharpest in \emph{multistage} workflows. We use \emph{multistage} to mean a project of two or more dependent calculations in which a downstream stage consumes an artifact (e.g., a converged charge density, a relaxed geometry, or a wavefunction) produced by an upstream stage; we call the transfer of that artifact a \emph{handoff}. Today's language models are already competent at a single calculation: given documentation, LLMs can usually select reasonable settings and write one valid input file. Their reliability collapses, however, once stages must be chained. Every handoff is a fresh opportunity to drop a dependency, point a stage at the wrong file, or ``invent'' an unsupported keyword, and these per-stage error rates compound multiplicatively.
Our direct-LLM runs make the failure modes concrete, as shown in Fig.~\ref{fig:llm_failure_modes}: the most common are incomplete multistage workflows with missing or unlinked stages, malformed structures, hallucinated input fields, and broken saved-density handoffs. A bare model that produces a runnable single calculation therefore still fails the chained, dependency-carrying workflows that production runs require. This gap is extremely consequential because materials discovery demands correctness by construction rather than by probability: a pipeline that succeeds ninety percent of the time is not nine-tenths as useful; across a campaign of thousands of calculations, it becomes a generator of silent errors and wasted solver time. Flagship LLM models may narrow this gap but not close it: they raise per-stage reliability yet still provide no guarantee, at a per-calculation cost that does not scale to high-throughput applications. To our knowledge, no existing first-principles agent delivers multistage reliability that is at once near-certain and inexpensive.

The errors (Fig.~\ref{fig:llm_failure_modes}) above share a decisive property: all of them are detectable from the project state before the solver ever runs, yet in expensive DFT workflows they are typically discovered only after execution, wasting computation and surfacing the mistake far from its origin, as a failed or silently wrong downstream run rather than as the setup error that caused it (Fig.~\ref{fig:workflow_comparison}(a)). Acting on this property means separating two kinds of post-execution iteration. Some is unavoidable: in a research-grade calculation, the appropriate $k$-point mesh or real-space grid spacing is not known in advance and must be found by running convergence calculations, checking the result, and refining the settings until the calculation is production-ready (Fig.~\ref{fig:workflow_comparison}(a)). But the same loop is often burdened by the preventable workflow errors catalogued above, which need never reach the solver. When an agent discovers these only after execution, the solver remains the first validator and HPC time is still spent. It is the purpose of this work to introduce an agent architecture, LatticeMind, that keeps solver-based convergence testing where it belongs but moves the preventable workflow checks before execution, as schematically shown in Fig.~\ref{fig:workflow_comparison}(b).

To be more concrete, we demonstrate and analyze the power of the LatticeMind agent using the DFT packages RESCU~\cite{MichaudRioux2016} and VASP~\cite{Kresse1996}. We use RESCU for most of the discussions below. As shown in Fig.~\ref{fig:workflow_comparison}(b), starting from a human scientific request in natural language, LatticeMind retrieves RESCU documentation and project context, plans the calculation stages, and writes a \emph{workflow contract} (its central abstraction): a structured specification of the calculation that it commits to before the DFT solver runs. It then generates the RESCU inputs and helper files, checks the project against the contract and deterministic solver rules before any computation is spent, repairs failed checks within a bounded number of repair-and-revalidate attempts, and proceeds to the approved execution and post-processing, leaving a traceable project directory of inputs, validation records, logs, outputs, and a report tied to those artifacts. Reliability is therefore built into the workflow by construction, as a property of an explicit, checkable specification rather than of the language model alone. Set against the multistage problem above, this makes the LatticeMind agent, to our knowledge, the first DFT agent to render chained workflows both production-reliable and cheaper than reactive repair.

Our study is organized around one question: can multistage first-principles workflows be made reliable by architecture and validation rather than by the scale of the language model? This work makes six contributions toward answering it. First, it evaluates first-principles agents at the artifact level: rather than scoring only the final scientific answer (e.g., a band gap), a successful run must produce a runnable project whose required intermediate files are checked, whose stages are correctly linked, and whose computed observables are correct within the predefined tolerance. Second, it introduces a validation-gated architecture in which workflow contracts, deterministic DFT input checks, bounded repair, approval gates, and traceability records control whether generated files may proceed to execution, so preventable setup and dependency errors can be repaired or stopped before the solver is launched. Third, it evaluates this architecture on RESCU (and VASP) benchmarks spanning the most common DFT workflow families, showing that three smaller-scale non-flagship backend LLMs (GPT-5.4-mini, Gemini 3 Flash Preview, and Claude Sonnet 4.6) each pass the reference checks for 100\% of the 44 workflows under the same scoring standard. This isolates the contribution of workflow architecture and the validation harness from that of model scale. Fourth, it uses an ablation study (eight architecture variants over a 23-case subset, three repeats each) to identify the reliability mechanisms: the workflow contract gives the largest single gain and deterministic validation before DFT execution is the decisive component. The same study compares against the conventional trial-and-repair model, which generates inputs and fixes them only after a failed DFT run and is the dominant design of current first-principles agents~\cite{Hu2026TritonDFT,Soleymanibrojeni2026GENIUS}; this baseline reaches only 71\% of cases. Fifth, it tests the same workflow controls beyond the benchmark, on a realistic multistage application whose research protocol is not represented in any benchmark case. Sixth, to show the approach is not specific to one code, it applies the same architecture to a different DFT code (VASP) on a 19-case benchmark subset spanning the same workflow families, reaching a 100\% reference-passing rate.

As an application example, we use LatticeMind to study a Rh$_6$ cocatalyst on \InGaN, a nitride photocatalyst system relevant to solar water splitting. The scientific question is how the local electronic structure near the Rh-covered surface changes with In concentration. For LatticeMind, this provides a realistic multistage workflow: the project requires alloy and slab setup, linked RESCU calculations, saved-density reuse, projected-density-of-states (PDOS) analysis, region definitions, and consistent post-processing. This tests whether validation-gated workflow construction can organize and check the files and dependencies needed for a production-style materials study. 

This paper is organized as follows. Section II presents the architecture of LatticeMind. Section III discusses the evaluation protocol and Section IV presents the LatticeMind benchmarks. Section V presents an application of LatticeMind to the investigation of III-nitride-based photocatalysts. Section VI presents further discussions of LatticeMind, and the last section is reserved for a short summary. 

\section{LatticeMind Architecture}

\begin{figure*}[t]
\centering
\makebox[\textwidth][c]{\includegraphics[width=\FigArchitectureWidth]{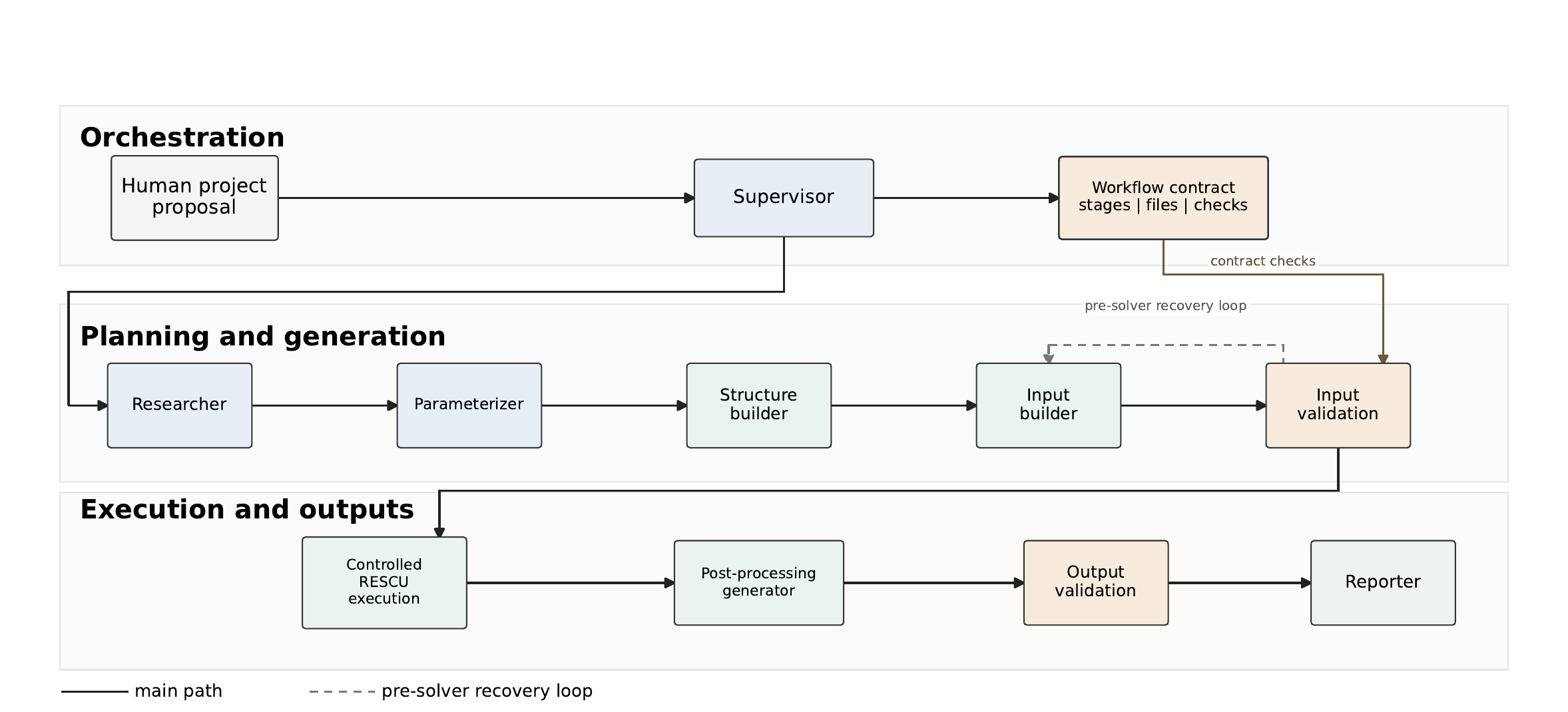}}
\caption{\label{fig:architecture}
LatticeMind architecture: seven specialized agents arranged in three color-coded layers. The human project proposal (top) enters the \emph{orchestration} layer, where a supervisor agent interprets the request and writes the workflow contract. The \emph{planning-and-generation} layer then carries out the plan: the researcher agent retrieves solver documentation and project context, the parameterizer agent selects calculation settings, the structure builder agent prepares geometries, and an input builder agent writes the input files for the DFT package, after which deterministic input validation checks the generated project before DFT execution. The \emph{execution-and-outputs} layer runs the approved input files through the DFT solvers and post-processing steps, and a reporter assembles a traceability report. Box colors indicate roles: blue boxes are reasoning agents (supervisor, researcher, parameterizer), green boxes generate or execute project files (structure builder, input builder, controlled RESCU execution, post-processing generator), orange boxes are the deterministic reliability mechanisms that constrain them (the workflow contract and the input and output validation), and neutral grey boxes mark the two endpoints of the run, the human proposal that enters the system and the reporter that returns the final report. Solid arrows show the main path. The single dashed (grey) arrow is the pre-solver recovery loop: a failed input-validation check is returned for targeted repair to the generation step responsible for the failed file and re-checked before any DFT solver runs; the arrow is drawn to the input builder, which handles the most common repairs, and a failed structure check is returned to the structure builder in the same way. This pre-solver recovery is the key component that lifts the full agent to a 100\% benchmark pass rate (Fig.~\ref{fig:ablation_summary}a), in contrast to the post-execution repair loop of Fig.~\ref{fig:workflow_comparison}(a).}
\end{figure*}

Figure~\ref{fig:architecture} shows the LatticeMind architecture. It has three main layers: Orchestration; Planning and generation; Execution and outputs. The remainder of this section explains what each part contributes, following the boxes of Fig.~\ref{fig:architecture} from the human project proposal at the top to the reporter at the bottom.

General language models are useful for interpreting scientific intent, but they are not naturally constrained by solver schemas, file-system state, or dependency rules: a model can write a valid self-consistent-field (SCF) to density-of-states (DOS) input, yet fail to link the two stages through the saved density. LatticeMind is designed to keep this language-level flexibility while adding the missing workflow constraints around it. The implementation evaluated here therefore uses language-model backends as provided, without RESCU (or VASP) specific fine-tuning. Solver specificity enters through retrieved documentation, the initial working directory, curated RESCU (or VASP) input rules, workflow contracts, deterministic validation, and targeted repair. LatticeMind is therefore not a RESCU- (or VASP-) trained language model, but a multi-agent workflow system in which deterministic validators and approval gates constrain when generated files may proceed to execution.

The workflow contract is written by the supervisor agent in the orchestration layer (the ``Workflow contract'' box in Fig.~\ref{fig:architecture}). It specifies, for a requested calculation, the required stages, the files each stage must produce or consume, the dependencies between stages, the execution action associated with each stage, and the checks that must pass before that action is allowed. These are exactly the procedural commitments that are usually left implicit in a script or discovered only at run time, and making them explicit and checkable in advance is what allows preventable errors to be caught while the project is still cheap to fix. It gives the validators concrete conditions to test before DFT execution. For example, in an SCF-to-DOS workflow, the DOS stage can be required to consume the density produced by the preceding SCF stage, rather than an unrelated or missing density file. The contract therefore makes stage-level validation possible before any costly solver step is launched.

Before the contract is written, the researcher agent in the planning-and-generation layer retrieves context from the documentation of the solver in use (e.g., RESCU or VASP) and from the initial working directory. The initial working directory is the directory supplied at the start of a run, before LatticeMind has generated new files. It may contain user-provided structures, template inputs, pseudopotentials, saved densities, or previous outputs allowed by the case definition. This retrieval step gives the agent access to solver-specific information available at run time and reduces reliance on model recall alone.

After the contract is written, the remaining boxes of the planning-and-generation layer carry it out: the parameterizer agent selects calculation settings, the structure builder agent prepares atomic-structure files, and the input builder agent writes the solver input files and helper scripts. Geometries that are complex (multicomponent systems such as alloys, slabs, or interfaces, rather than a documented bulk unit cell) or unsourced (constructed by the agent rather than taken from a user-supplied file or a retrieved reference structure) are additionally screened by an adversarial review before acceptance. Here, helper scripts are small post-processing programs used to parse solver outputs, extract quantities such as DOS or band paths, assemble tables, or produce plots. They are part of the project record because the final reported quantities should remain traceable to the files and commands that produced them.

Generated inputs for the solver in use (e.g., RESCU or VASP) then reach the ``Input validation'' box of Fig.~\ref{fig:architecture}, which is the validation gate of the architecture: the inputs are checked by deterministic validators rather than by a second language-model judgment. These validators check solver-specific requirements assembled from the solver's documentation, tutorial inputs, and tested solver conventions, including:
\begin{itemize}
\item the set of input fields that may be written, the expected value type of each field, and known aliases (which are converted to solver-specific fields);
\item the required fields for each calculation type;
\item pseudopotential path requirements;
\item the stage dependencies declared in the workflow contract (e.g., that a DOS stage consumes the saved density of its SCF stage);
\item band-structure path conventions; and
\item simple geometric constraints on the generated structure.
\end{itemize}
If a check fails, LatticeMind sends the specific validation error back to the generation step responsible for the failed file, along the dashed grey arrow of Fig.~\ref{fig:architecture}; this pre-execution loop is the validation-gated repair loop referred to throughout this paper. Repair is attempted only up to a fixed number of repair-and-revalidate cycles, and if the required checks still do not pass, the workflow stops rather than claiming completion without checked files. This makes the validation gate explicit: the DFT execution proceeds only after the generated project satisfies the contract and the relevant input rules.

The validated project then enters the execution-and-outputs layer, where execution-sensitive steps are controlled by human approval gates. In an interactive run, the user can approve structure generation, solver execution, and post-processing before LatticeMind proceeds. In the benchmark setting used here, the same approval policy was fixed across all cases, and low-cost solver and post-processing steps were allowed to run automatically when permitted by the case definition. In both settings, the purpose of approval is to prevent uncontrolled execution: an action is allowed only after the required checks have passed and the relevant approval condition is satisfied.

Generated helper scripts are run in a restricted environment. This is a safety constraint for generated analysis code, not an assumption that the agent normally attempts unsafe actions. The restricted environment blocks network access and subprocess calls and limits file writes to approved project locations. This prevents generated analysis code from changing files outside the project record and helps preserve the traceability of the run.

The layer ends with the reporter agent, which assembles the project record into a polished, human-readable report that remains tied to the generated files and validation records behind it.

\section{Evaluation Protocol}

\subsection{Benchmark design and scoring}

The \emph{LatticeMind Benchmark for RESCU} evaluates whether LatticeMind produces a usable project under a predefined task specification for the RESCU DFT package~\cite{MichaudRioux2016}. We report three primary outcomes. \emph{Output produced} means that the expected solver output files or post-processing artifacts are present. \emph{Workflow complete} means that the requested calculation stages and required handoffs between stages are present and internally consistent. \emph{Passed checks} means that the completed workflow produces the requested scientific result within the predefined tolerance for that case, while also satisfying any required file, dependency, or input-value checks. For example, in an SCF-to-DOS task, output production requires that both SCF and DOS result files exist; workflow completion requires that the two stages be linked through the saved density; and passing checks requires that the DOS-derived band gap or another requested observable fall within the predefined tolerance. A generated project that looks plausible in text but lacks a required dependency, contains unsupported input fields, or reports an unchecked output is therefore not counted as successful, because it would not be safe to scale into an automated DFT campaign. The benchmark does not, however, require a single arbitrary file layout or one exact implementation when the prompt leaves those choices open: two projects may both pass with different harmless temporary file names, plotting styles, or intermediate script names, provided they use the required calculation stages, preserve the required file handoffs, and produce the requested observable. The reference physical values used for these scientific-output checks were drawn from Materials Project entries~\cite{Jain2013MaterialsProject} or from the literature, and the full scoring rubric is given in the Supplemental Material (Sec.~S4).

\subsection{RESCU benchmarks}

The benchmark suite contains 44 cases spanning total-energy, density-of-states, projected-density-of-states, band-structure, magnetic, phonon, potential, relaxation, and band-edge workflows; the per-family composition and the full case inventory are given in the Supplemental Material (Sec.~S3). Prompts were constructed by modifying RESCU tutorials, documentation examples, and representative workflow templates: every case was changed in material system, crystal prototype, calculation chain, or requested observable, so no benchmark case is identical to any published tutorial or documentation example, and none is answerable by reproducing one. The near-zero pass rate of the direct-LLM baselines on these same cases (Fig.~\ref{fig:ablation_summary}) confirms this empirically: if the benchmark were solvable from memorized examples, a bare model would not fail almost every case. A companion 19-case VASP benchmark, built from the same workflow families to test solver transfer, is described together with its results in Sec.~\ref{sec:vasp_transfer}.

\begin{figure*}[t]
\centering
\makebox[\textwidth][c]{\includegraphics[width=\FigExamplePromptWidth]{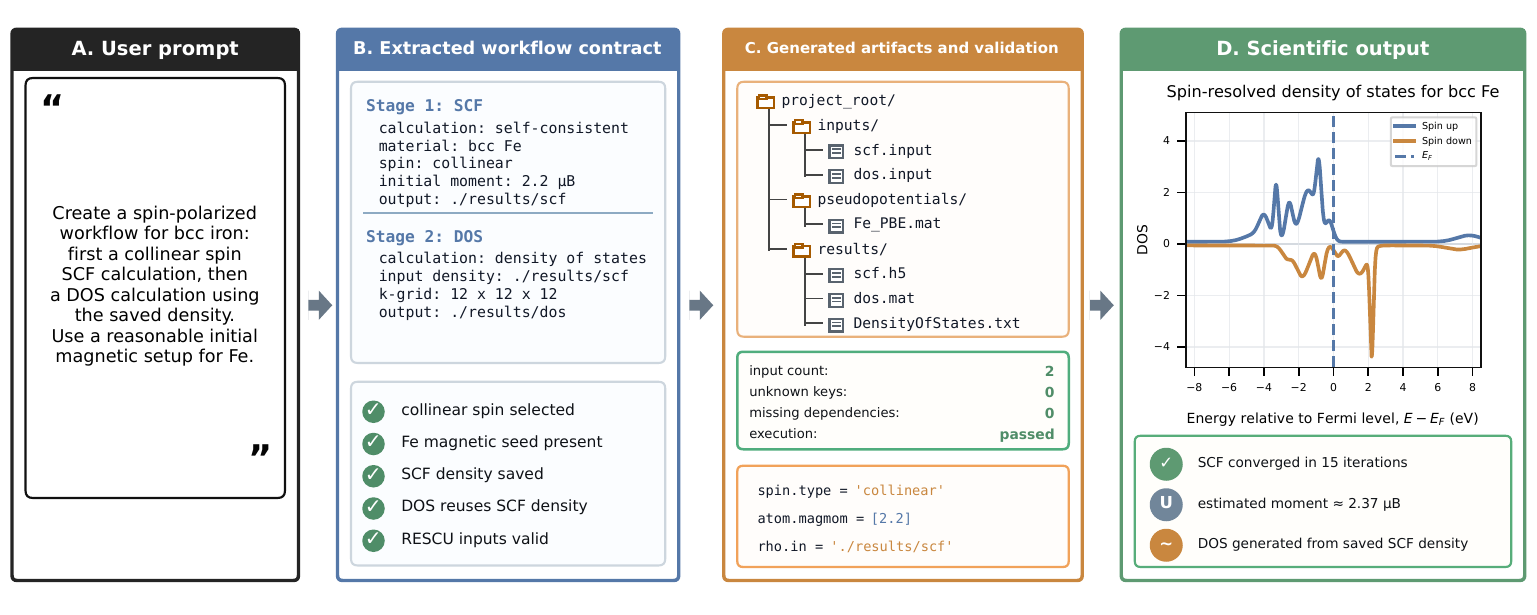}}
\caption{\label{fig:example_prompt_workflow}
Representative prompt-to-project workflow for a bcc Fe spin-polarized SCF-to-DOS benchmark case. The panels are a schematic built from representative LatticeMind outputs, not screenshots of a LatticeMind interface. (a) The user prompt specifies the material, spin treatment, workflow order, and saved-density requirement in natural language. (b) LatticeMind extracts a workflow contract with the required stages, saved-density link, and checks. (c) The generated RESCU project is evaluated through concrete files and validation checks, including SCF and DOS inputs, pseudopotentials, result files, and checked input values. (d) The scientific output remains tied to the generated project; in this example the DOS calculation uses the saved SCF density and produces the spin-resolved DOS.}
\end{figure*}

Before the benchmark, we froze three items for each case: the prompt, the initial working directory, and the scoring specification. The scoring specification told the external scorer what to look for in the finished project, such as the required calculation stages, file handoffs, output artifacts, and scientific-output checks. It was not given to LatticeMind as an answer key. During the run, LatticeMind still had to generate its own workflow contract, RESCU inputs, dependency links, execution records, and output files. We did not revise prompts, starting files, or scoring criteria after seeing generated outputs. This prevents \emph{post hoc} prompt tuning and makes the systems comparable. Figure~\ref{fig:example_prompt_workflow} traces one representative case through the architecture of Fig.~\ref{fig:architecture}: panel (a) is the human project proposal that enters the orchestration layer; panel (b) is the workflow contract the supervisor agent commits to; panel (c) is the generated RESCU project, the input files written by the input builder and screened by deterministic input validation; and panel (d) is the scientific output, which the reporter keeps tied to the checked project.

\subsection{Systems, execution protocol, and scoring}

The full benchmark used the identical LatticeMind architecture across three model backends, differing only in the language model and its decoding settings (Table~\ref{tab:backends}). Exact run dates and the full system manifest are reported in the Supplemental Material (Table~S1).

\begin{table}[t]
\caption{\label{tab:backends}Model backends for the LatticeMind Benchmark for RESCU. All three run inside the same LatticeMind configuration (local RESCU documentation, a workflow contract, deterministic pre-execution validation, at most three validation-triggered repair cycles, and no web search) and are graded by the same external scoring script; only the backend and its decoding settings differ. Decoding settings: GPT-5.4-mini, reasoning effort ``quick''; Gemini 3 Flash Preview, $T=0.1$ and top-$p=0.5$; Claude Sonnet 4.6, provider defaults.}
\small
\begin{ruledtabular}
\begin{tabular}{@{}ll@{}}
Backend & Model identifier \\
\colrule
GPT-5.4-mini & \texttt{gpt-5.4-mini} \\
Gemini 3 Flash Preview & \texttt{gemini-3-flash-preview} \\
Claude Sonnet 4.6 & \texttt{claude-sonnet-4-6} \\
\end{tabular}
\end{ruledtabular}
\end{table}

Each run begins from a clean working directory containing only the files allowed by the case definition, and the same benchmark scripts, starting-directory rules, approval settings, execution policy, and scoring checks were used for every system. During a run, LatticeMind generates and validates project files, performs approved execution steps (low-cost RESCU and post-processing steps run automatically when permitted by the case definition), and stores inputs, logs, outputs, and scoring records; no case was manually corrected before scoring. All run directories were then processed by an automatic scoring script that operated outside the agent loop and did not call a language model, and each case-system pair was run once; the statistical scope of this choice is discussed below. Detailed task manifests, scoring definitions, traceability records, ablation baselines, and generated benchmark tables are provided in the Supplemental Material (Secs.~S6 and~S8).

\subsection{Matched architecture ablation}

We evaluate eight architecture variants on a 23-case subset of the benchmark (GPT-5.4-mini, 52\% of the 44 cases), with three repeats per case. The variants comprise the full LatticeMind agent; four single-component ablations, each removing one reliability mechanism (documentation retrieval, the workflow contract, the agent-side validator, or validator-triggered repair); a trial-and-repair baseline (the dominant design of current first-principles agents~\cite{Hu2026TritonDFT,Soleymanibrojeni2026GENIUS}) that generates inputs and then repairs them only after a failed solver run, with no validation beforehand; and two direct-LLM baselines (with and without documentation) that use the model alone. The first five variants support a leave-one-out reading, and the full set supports a cumulative build-up reading in which components are added one at a time, from the language model alone to the full agent. Every variant keeps the prompts, run settings, external deterministic scoring script, and predefined checks fixed, so differences isolate component effects; three repeats per case separate transient execution noise from genuine failures. For example, removing the workflow contract tests whether the system can still preserve required stage dependencies, such as an SCF-to-DOS saved-density handoff; removing the validator tests whether generated RESCU inputs remain valid without deterministic field and dependency checks.

\section{Benchmark results}

This section reports the RESCU benchmark, the matched architecture ablation, and the VASP transfer test; the \RhInGaN\ photocatalyst application follows in Sec.~\ref{sec:application}.

\begin{figure*}[tbp]
\centering
\includegraphics[width=\FigBenchmarkWidth]{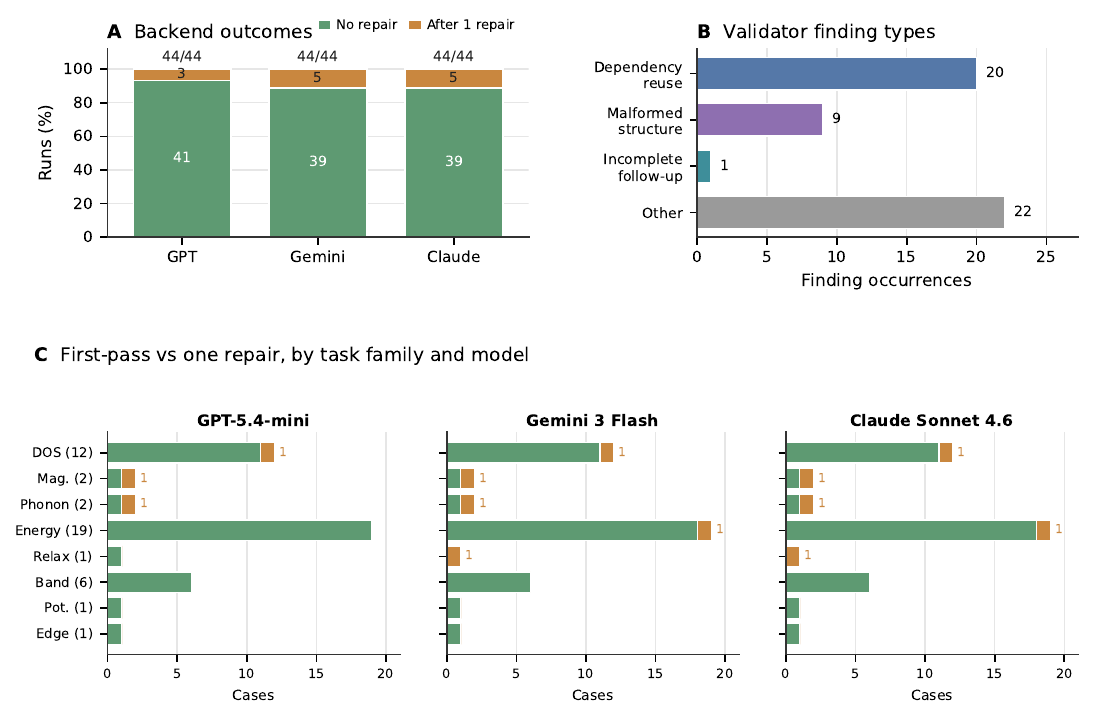}
\caption{\label{fig:benchmark_summary}
RESCU benchmark results from the completed run directories, for three non-flagship backends: GPT-5.4-mini (GPT), Gemini 3 Flash Preview (Gemini), and Claude Sonnet 4.6 (Claude). All three backends produced outputs, completed workflows, and passed reference checks in all 44 cases. Panel (a) separates workflows that passed pre-execution validation without repair from those that passed after one validation-gated repair cycle; no runs remained unrecovered. Panel (b) summarizes validator-finding occurrences logged during pre-execution checks; categories may co-occur in one run. Panel (c) aggregates all 132 runs by task family, with run counts in parentheses, using the same no-repair and one-repair colors as panel (a).}
\end{figure*}

Figure~\ref{fig:benchmark_summary} summarizes the 44-case benchmark. GPT-5.4-mini, Gemini 3 Flash Preview, and Claude Sonnet 4.6 each produced complete workflows and reproduced all 44 benchmark reference values within tolerance under the same scoring standard. The endpoint was therefore identical across the three evaluated backends; the remaining differences are in how often the validation-gated repair loop was needed before execution readiness.

Because every case ultimately passed, the main distinction is whether pre-execution repair was needed (Fig.~\ref{fig:benchmark_summary}a,c). Across all 132 runs, 119 (90\%) passed on the first validated plan and 13 (10\%) passed after one validation-gated repair cycle (the green loop in Fig.~\ref{fig:workflow_comparison}(b)); no run required more than one repair cycle. GPT-5.4-mini required validation-gated repair in 3 cases, while Gemini 3 Flash Preview and Claude Sonnet 4.6 required repair in 5 cases each. These validation-gated repairs concentrated in five demanding cases (magnetic NiO, the finite-displacement Si phonon, the ordered-alloy SiGe relaxation and its DOS follow-up, and the trigonal $\alpha$-quartz), rather than in any whole task family. Every other case, including all band-structure, potential, and band-edge workflows, passed on the first validated plan. The validator-finding taxonomy (Fig.~\ref{fig:benchmark_summary}b) shows that dependency reuse, malformed structure or lattice information, and incomplete follow-up workflows were caught before DFT execution, with a residual ``other'' category for less frequent findings. Cleaning generated inputs against the allowed RESCU input fields and rules increased the fraction of schema-valid inputs from 37.9\% to 84.1\% (pooled across the three backends). These are precisely the kinds of errors that should be found before an automated workflow is allowed to consume solver or HPC resources.

\subsection{Matched architecture ablation}

\begin{figure*}[!t]
\centering
\makebox[\textwidth][c]{\includegraphics[width=\FigAblationWidth]{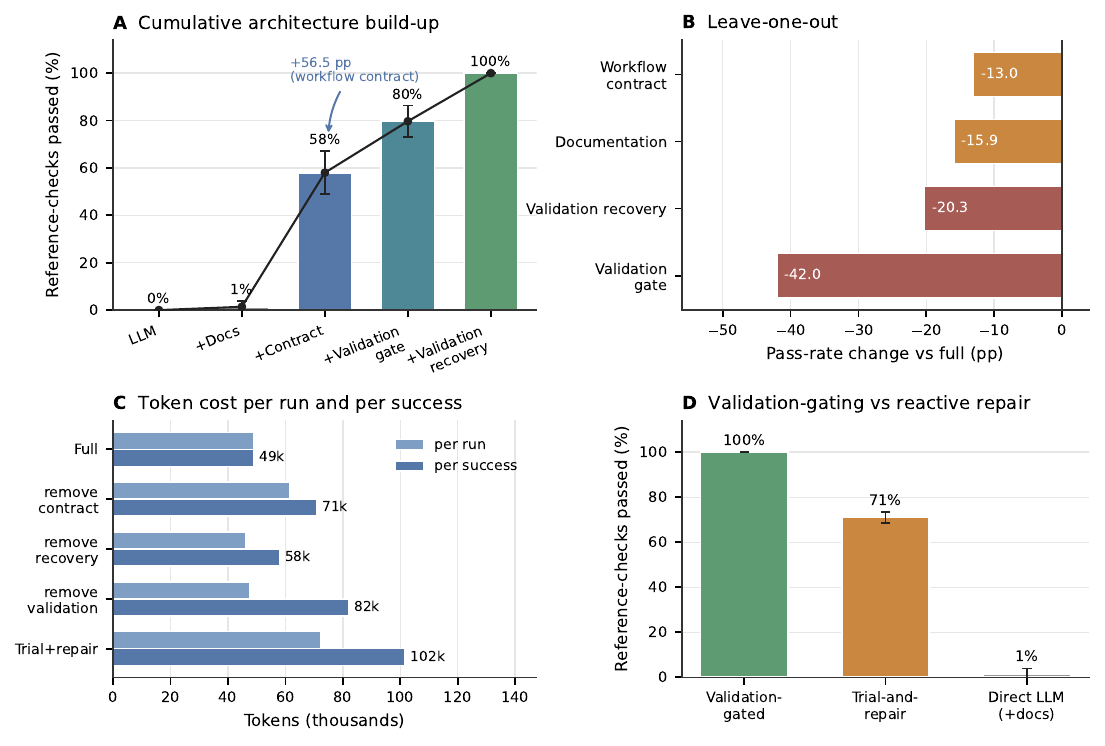}}
\caption{\label{fig:ablation_summary}
Architecture ablation (eight variants, 23-case subset, three repeats per case). (a) Cumulative build-up: starting from the language model alone, each component is added in turn (documentation, then the workflow contract, then the validation gate, then pre-execution validation recovery $=$ full LatticeMind); the workflow contract is the largest single gain ($+56.5$ percentage points). Error bars in the pass-rate panels give $\pm1$ standard deviation of the pass rate across the three repeat rounds (run-to-run variability). (b) Leave-one-out: change in reference-check pass rate when one component is removed from the full agent. (c) Token cost per run and per successful result. (d) Validation-gated generation versus a reactive trial-and-repair baseline and a direct LLM with documentation.}
\end{figure*}

Figure~\ref{fig:ablation_summary} summarizes the matched ablation (eight variants $\times$ 23 cases $\times$ 3 repeats $=$ 552 runs), which both builds the architecture up one component at a time and removes components one at a time. All variants share the same model, scoring policy, and predefined checks, so the differences estimate component effects under a controlled comparison.

Built up from the LLM alone, reliability rises component by component (Fig.~\ref{fig:ablation_summary}a). Surprisingly, a direct LLM passes 0\% of cases. Adding documentation reaches 1.4\%; adding the workflow contract (the ``Workflow contract'' box in Fig.~\ref{fig:architecture}) reaches 58.0\%; adding the deterministic validation gate (the ``Input validation'' box, which at this build-up step only rejects projects that fail its checks) reaches 79.7\%; and adding pre-execution validation recovery (the dashed repair loop that returns a failed check for correction and revalidation), completing the full LatticeMind agent, reaches 100\%. The workflow contract is the single largest gain, a 56.5 percentage-point jump, consistent with its role as the central abstraction: it is what first turns a fluent but unconstrained model into a system that produces dependency-consistent, multistage projects. The size of this jump follows from the failure modes of Fig.~\ref{fig:llm_failure_modes}: a bare model most often fails by omitting or unlinking whole stages, and the contract removes exactly this error class by forcing the required stages, files, and handoffs to be enumerated before any input file is written, leaving the downstream gate and repair loop to catch the remaining field-level and dependency errors.

The final step is \emph{validation recovery}: the inner, pre-solver loop in which a failed validator check is sent back for repair and re-checked before any DFT execution (the grey loop in Fig.~\ref{fig:architecture}). It is distinct from the post-solver \emph{replan recovery} of the trial-and-repair baseline below, in which a failure is diagnosed only after the DFT solver has run. The two recoveries behave very differently. In the full 44-case benchmark, validation recovery fired on 13 of 132 runs, while the post-execution replan loop fired on none: the validation gate caught every preventable error before the solver ran. The size of the recovery step should therefore be read relative to the case set: the ablation subset is deliberately harder than the full benchmark (52\% of the cases, weighted toward multistage families), so recovery contributes about 20 points here, versus the 10 points it adds on the full benchmark.

Removing one component at a time from the full agent (Fig.~\ref{fig:ablation_summary}b) shows that the validation gate (the ``Input validation'' box in Fig.~\ref{fig:architecture}) is the most critical single component: removing it drops the full system from 100\% to 58.0\%. Because the recovery loop acts only on validator findings, removing the gate disables recovery with it, so this arm measures the two mechanisms together; consistently, it lands at the same 58.0\% as the documentation-plus-contract stage of the build-up. Removing the workflow contract alone, by contrast, lowers the pass rate less, to 87.0\%, while raising token cost by about 26\% per run, because the validation gate and recovery loop catch and repair many of the same errors the contract would otherwise have prevented. The components are therefore deliberately redundant, but not equivalent: documentation, the contract, and recovery partially compensate for one another, whereas removing the validation gate also disables recovery and produces the largest leave-one-out loss.

This redundancy is exactly what the conventional alternative agent lacks. A trial-and-repair baseline (Fig.~\ref{fig:workflow_comparison}(a)), generation followed by reactive post-execution repair with no pre-execution validation, reaches only 71.0\% (Fig.~\ref{fig:ablation_summary}d), and this post-execution replan recovery fired on 14 failed runs but corrected only 5: post-hoc repair cannot recover the structural errors, such as missing stages or absent input decks, that a pre-execution contract prevents. It is also the most expensive variant per successful result, at roughly 102k tokens per pass versus 49k for the full agent (Fig.~\ref{fig:ablation_summary}c). In short, the workflow contract lowers token cost while the validation gate raises the success rate: validation-gated generation is both more reliable and cheaper per usable result than post-execution reactive repair.

\subsection{Solver-agnostic generalization}\label{sec:vasp_transfer}

The architecture of LatticeMind is very general and can be applied to any other DFT package. To demonstrate this, we applied it to a different DFT code, the Vienna \textit{Ab initio} Simulation Package (VASP)~\cite{Kresse1996}, on a 19-case benchmark subset spanning the same workflow families as the RESCU benchmark presented above and weighted toward the multistage chains. The parts that are specific to RESCU (the allowed input fields and validation rules, the documentation source, and the helper scripts) enter only through configuration, so adapting to VASP required changing that configuration, not the workflow logic. Using the same GPT-5.4-mini backend, LatticeMind reached a 100\% reference-passing rate (19 of 19 cases), with the converged lattice constants and band gaps of the multistage cases falling within Materials Project and PBE reference windows. Per-case results are listed in the Supplemental Material.

VASP is also the most widely used DFT code, so an LLM has likely seen far more VASP input than RESCU input, raising the question of whether the architecture still helps when the backend is most fluent. To check this, we ran a direct-LLM baseline (the same model asked in one shot to write the complete VASP inputs, with no contract, validation, or repair) on the same 19 cases and scored each on the lenient bar of whether the generated inputs run to completion. The baseline produced runnable VASP inputs in about a third of the cases (7 of 19), far above the at-most-1.4\% it reached on RESCU, confirming that more training exposure raises the baseline competence of LLMs. The failures are systematic: the baseline handles self-contained single calculations but breaks down on the multistage workflows that reuse a converged charge density. In none of the 13 dependency chains does it construct the cross-stage density handoff that the workflow contract makes explicit. Stronger raw familiarity with a code therefore raises the floor on individual calculations but does not, by itself, yield correct multistage workflows, and LatticeMind still reaches 100\% on both codes (Supplemental Material, Sec.~S9 and Fig.~S2).

\section{Application of LatticeMind: \texorpdfstring{Rh$_6$/In$_x$Ga$_{1-x}$N photocatalyst}{Rh6/InxGa1-xN photocatalyst}}
\label{sec:application}

The benchmarks above evaluate LatticeMind on frozen, predefined tasks. This section applies it to a production research problem: how a Rh$_6$ cocatalyst modifies the local electronic structure of the \InGaN\ photocatalyst surface as the In concentration varies. The emphasis throughout is on the question the benchmarks leave open: whether LatticeMind's validated workflow construction extends to a research protocol unlike any it was benchmarked on (Sec.~\ref{sec:workflow_construction}). The photocatalyst physics is the test bed that makes this question concrete; the goal is not to establish a new photocatalytic mechanism, and the detailed region- and layer-resolved methodology that supports the physical reading is collected in the Supplemental Material (Sec.~S10).

\subsection{Scientific motivation and production protocol}\label{sec:app_protocol}

The \RhInGaN\ application was chosen because it represents a realistic research workflow in which meaningful conclusions require coordinating multiple dependent DFT stages: interface construction, chained electronic-structure calculations, region-resolved projected analysis, and consistent post-processing. Beyond the single-answer benchmark tasks above, it tests whether LatticeMind can manage the complexity and traceability required in a production-level materials study. The underlying science makes these demands concrete. Rh-decorated III-nitride photocatalysts combine solar-fuel relevance with local interface physics that is difficult to capture with a single laterally averaged descriptor. Photocatalytic water splitting requires efficient solar-light absorption, suitable band-edge energetics, effective charge separation and transfer, active surface redox sites, and long-term chemical stability~\cite{Li2017,Hisatomi2019}. \InGaN\ is attractive because its alloy composition tunes electronic structure and band-edge positions while retaining the chemical robustness of nitride semiconductors~\cite{Vurgaftman2001,Wang2019,Zhou2023}, and recent high-efficiency InGaN-based water-splitting results make this material class a serious target for first-principles interface analysis~\cite{Zhou2023}. A Rh cocatalyst can help extract carriers and promote surface reactions, but it can also reshape near-gap \InGaN\ states through local band bending, hybridized metal states, metal-induced gap states, and lateral electrostatic heterogeneity~\cite{Kamat2012,Tersoff1984,Liu2024}. Following region-resolved work showing that the nitride region beneath a nanocluster and the uncovered nitride surface can play different electronic roles~\cite{Yuan2026}, we ask how a finite Rh$_6$ cocatalyst modifies the local near-gap states at the Rh-covered \InGaN\ surface, and how that response changes with In content and local In/Ga configuration.

Methodologically, the application exercises the full LatticeMind workflow on a multistage DFT problem rather than on isolated benchmark tasks, and its production operations entered the framework the same way the benchmark case of Fig.~\ref{fig:example_prompt_workflow} does. Each operation began as a natural-language request naming the structure (a host slab, its Rh$_6$-decorated counterpart, or a replicate unit), the calculation chain, and the projected quantities required from it. The supervisor translated each request into a workflow contract enumerating the stages, the files every stage consumes and produces, and the checks that gate execution, and the generated RESCU inputs had to pass the deterministic validators of Sec.~II before any solver time was spent. For this protocol, the contract carries in particular the saved-density handoffs that link each self-consistent slab calculation to its projected-DOS stages, and the explicit atom, orbital, layer, and region assignments that define the region-resolved projections introduced below; both are checked before execution (Supplemental Material, Sec.~S10). The study spans a series of special-quasirandom-structure (SQS) alloy compositions and, for each, this chain of dependent electronic-structure, projection, and post-processing stages. The high-level scientific protocol was expert-guided; LatticeMind constructed and validated the projects that realize it, and Sec.~\ref{sec:workflow_construction} quantifies that construction.

Following this LatticeMind-assisted protocol, the photocatalyst study uses bulk wurtzite In$_x$Ga$_{1-x}$N alloys as the starting point. For each alloyed composition, $x=0.125$, $0.25$, and $0.375$, we use the SQS approach~\cite{Zunger1990} to generate three independent SQS configurations; the corresponding wurtzite parent cells provide the bulk references for the slabs. We also include an $x=0$ GaN calculation point for reference.

From each bulk alloy reference, we cleave a nonpolar (110) sidewall slab, decorate it with a six-atom Rh nanocluster placed on the nitride-terminated surface, and relax it at fixed cell parameters. Geometry relaxation of the decorated slab is performed using the MACE-MPA-0 machine-learning interatomic potential with the \texttt{medium-mpa-0} checkpoint~\cite{Batatia2025}. The final electronic-structure calculation on each relaxed geometry uses the Perdew--Burke--Ernzerhof (PBE) functional~\cite{Perdew1996}, as implemented in the real-space RESCU code~\cite{MichaudRioux2016}. We did not search globally over \RhInGaN\ interface motifs. The electronic results apply to the relaxed motifs studied here.

With the structural model fixed, we specify the numerical settings used to compute the electronic structure of each relaxed slab. Each relaxed \RhInGaN\ slab contains 582 atoms and is surrounded by at least 22~\AA\ of vacuum along the slab-normal direction. Brillouin-zone sampling uses a $\Gamma$-point-only $1\times1\times1$ grid, which is consistent with the large in-plane supercell. The real-space grid resolution was set to 0.14~bohr. Neumann electrostatic boundary conditions are imposed in the slab-normal direction, while the in-plane directions remain periodic. Density-of-states output is broadened with a Gaussian of width 0.05~eV.

Because Rh$_6$ covers only part of the nitride surface, the covered and uncovered parts of the slab are distinct local environments. A laterally averaged DOS would mix the substrate atoms directly beneath the cluster with substrate atoms on the exposed surface. Following the region-resolved interface picture of Yuan \emph{et al.}~\cite{Yuan2026}, we therefore analyze these two environments separately.

The \InGaN\ substrate atoms are first grouped into eight layers along the slab-normal direction. The layer index increases from the lower side of the slab to the Rh-covered side, so that $L=1$ is the lower substrate layer and $L=8$ is the top \InGaN\ layer adjacent to Rh$_6$. Within every layer we then define two lateral regions from the projected Rh$_6$ footprint. Region A is the set of \InGaN\ substrate atoms whose in-plane positions fall inside the Rh$_6$ footprint after adding a $1.50$~\AA\ margin. Region B is the complementary set of \InGaN\ substrate atoms in the same layer and in the same decorated-slab calculation. Thus, Region B represents the uncovered part of the decorated slab; it is not a separate bare-slab calculation and is not rescaled by trimming atoms or matching area to Region A. Comparing Regions A and B within the same decorated slab keeps the alloy configuration, slab geometry, and electrostatic environment fixed while isolating the local perturbation associated with the Rh footprint. In the generated projects, these region and layer assignments are declared explicitly in each projected-DOS input rather than applied afterward as a plotting convention.

The next step is to make the Region-A and Region-B spectra comparable. Region B covers a larger lateral area than Region A, so it usually contains more projected atoms, and comparing raw PDOS integrals would make Region B look larger simply because more atoms were included. We therefore divide each Region-A or Region-B PDOS curve by the number of selected \InGaN\ substrate atoms in that region and layer, converting the comparison from total spectral weight over differently sized atom lists to spectral weight per selected substrate atom. After this normalization, the substrate PDOS is reported in units of states/eV/atom, and an energy integral has units of states/atom. Rh atoms are not included in these substrate-region spectra, because the purpose of the Region-A/Region-B comparison is to ask how the \InGaN\ substrate changes beneath the nanocluster; the Rh$_6$ cluster is projected separately and used only for the Rh-orbital metric below. Throughout, \emph{PDOS} refers to a projected density of states for a specified atom list, species, or orbital channel, and \emph{PLDOS} (projected local density of states) to the same information organized layer by layer, following the region-resolved PLDOS language of Yuan \emph{et al.}~\cite{Yuan2026}.

We then define a consistent energy window that represents the \InGaN\ gap for each alloy configuration. The gap is defined from Region B, the uncovered part of the same decorated slab. This choice follows Yuan \emph{et al.}'s physical separation of Region A, where nanocluster-induced band bending and gap states are expected, from Region B, where the uncovered nitride surface retains its own surface-state character~\cite{Yuan2026}. Using Region B also keeps the reference within the same SQS configuration and electrostatic environment as Region A, rather than comparing to a separate bare-slab calculation.

For each SQS configuration, the fixed post-processing protocol searches the interior Region-B layer spectra for a clean gap. In the eight-layer slabs used here, the candidate Region-B layers are $L=3$--6, which avoids both outer surfaces. The valence onset is read from the N-projected PDOS, and the conduction onset is read from the Ga+In cation-projected PDOS\@. This assignment follows the usual orbital character of III-nitride band edges and the same physical idea used by Yuan \emph{et al.}, who identify N-derived and Ga-derived band-edge markers from local PLDOS~\cite{Yuan2026}. Our implementation uses thresholded onsets so that the same automated rule can be applied to every sampled structure. The threshold is swept over a small set of values; windows with implausible gap widths are rejected, and the accepted window with the least residual substrate PDOS inside the gap is chosen. The threshold list, allowed gap range, and scoring expression are given in the Supplemental Material (Sec.~S10\,C).

The chosen Region-B spectrum defines two reference energies, $E_V^{\mathrm{ref}}$ and $E_C^{\mathrm{ref}}$. Their difference is $\Egref = E_C^{\mathrm{ref}} - E_V^{\mathrm{ref}}$. The interval $[E_V^{\mathrm{ref}},E_C^{\mathrm{ref}}]$ is then used as the gap window for the integrated quantities for that SQS configuration. We denote by $L_{\mathrm{gap}}$ the Region-B layer that supplies this gap window. This layer is only used to define the energy window; it is distinct from the interior Region-A layer $L_{\mathrm{ref}}$ used below to compare band-edge positions.

If no clean projected Region-B gap is found under this rule, we treat that outcome as part of the spectra rather than as a failed calculation. Gap-window quantities such as $\Egref$, $\DEC$, and $\DNgap$ are then not assigned for that SQS configuration. For example, if the projected Region-B spectrum has residual states throughout the expected gap region, assigning a valence and conduction onset would be more arbitrary than informative. We instead use fixed near-Fermi descriptors that do not require identifying a clean gap window, so these configurations can still be compared without forcing a band-edge interpretation. In particular, we report
\begin{equation}
\DNfermi =
\int_{-0.1\,\mathrm{eV}}^{+0.1\,\mathrm{eV}}
\left[\, \rho_A(E,L_{\mathrm{surf}}) - \rho_B(E,L_{\mathrm{surf}}) \,\right]\,dE,
\end{equation}
where energies are measured relative to the Fermi level and the PDOS curves are normalized per selected substrate atom as described above. This quantity asks whether the covered nitride surface has excess substrate-projected weight near the Fermi level without forcing a band-edge interpretation onto spectra that do not contain a clean projected gap.

With the regions and reference gap defined, we compute three gap-window descriptors for each gap-defined configuration. The first descriptor tests whether the Region-A conduction edge is lower at the Rh-covered surface than in the Region-A interior. For our slab convention, we define
\begin{equation}
\DEC = E_C^{A}(L_{\mathrm{ref}}) - E_C^{A}(L_{\mathrm{surf}}),
\end{equation}
where $E_C^{A}(L)$ is the Region-A conduction-band onset at layer $L$, determined by the same onset rule used above. The surface layer is $L_{\mathrm{surf}}=8$, the top \InGaN\ layer under the Rh cluster. The interior comparison layer is $L_{\mathrm{ref}}=5$, and sensitivity to nearby interior choices is reported in the Supplemental Material (Sec.~S10\,D). A positive $\DEC$ means that the conduction onset is lower at the Rh-covered surface than in the Region-A interior, which is the sign expected for downward electron-collecting band bending. Because the number depends on the chosen onset threshold and reference layer, we report $\DEC$ as a fixed-protocol descriptor, not as a threshold-independent electrostatic band-bending measurement.

The second descriptor measures whether the Rh-covered \InGaN\ surface gains extra spectral weight inside the Region-B gap. This follows the region-resolved idea of quantifying nanocluster-induced gap states from the covered-region PLDOS~\cite{Yuan2026}, but we subtract the uncovered surface contribution so that ordinary Region-B surface states are not counted as Rh-induced excess:
\begin{equation}
\DNgap = \int_{E_V^{\mathrm{ref}}}^{E_C^{\mathrm{ref}}} \left[\, \rho_A(E, L_{\mathrm{surf}}) - \rho_B(E, L_{\mathrm{surf}}) \,\right]\,dE,
\end{equation}
where $\rho_A(E,L_{\mathrm{surf}})$ and $\rho_B(E,L_{\mathrm{surf}})$ are the per-substrate-atom PDOS curves for the top \InGaN\ layer in Regions A and B. The result is a covered-minus-uncovered gap-window DOS difference in states per substrate atom. A positive $\DNgap$ means that the \InGaN\ atoms directly under Rh$_6$ have more substrate-projected near-gap spectral weight than the uncovered atoms in the same layer of the same slab. This quantity is also related to metal-induced gap states at extended metal--semiconductor interfaces~\cite{Tersoff1984,Kamat2012,Liu2024}, but here the perturbation is spatially confined to the finite Rh$_6$ footprint.

The third descriptor reports the orbital character of the Rh spectral weight inside the same gap window. We quantify the Rh $d$ fraction as
\begin{equation}
f_{d}^{\mathrm{Rh}} = \frac{\int_{E_V^{\mathrm{ref}}}^{E_C^{\mathrm{ref}}} \rho_{\mathrm{Rh},d}(E)\,dE}{\int_{E_V^{\mathrm{ref}}}^{E_C^{\mathrm{ref}}} \rho_{\mathrm{Rh}}(E)\,dE},
\end{equation}
where $\rho_{\mathrm{Rh},d}$ is the Rh $d$-projected cluster PDOS and $\rho_{\mathrm{Rh}}$ is the total Rh-cluster PDOS over the projected Rh $s$, $p$, and $d$ channels. Both integrals use the Region-B reference gap window.

\subsection{Workflow construction for a novel research protocol}\label{sec:workflow_construction}

The benchmarks above establish reliability on workflow shapes drawn from the standard vocabulary of DFT practice: documented calculation types, each with a reference answer to check against. Research problems, by contrast, are defined by protocols that appear in no tutorial and for which no reference workflow exists. Whether validated construction extends from these familiar task shapes to a genuinely new protocol is a question the benchmarks themselves cannot answer, and it is the one this application addresses.

No single documented RESCU calculation corresponds to the workflow this study requires. Each individual component is standard: relaxations, self-consistent slab calculations, and projected-DOS analyses of the kinds the benchmark already covers. Their composition is not. The protocol above, which compares the Rh-covered and uncovered parts of one decorated slab through per-layer projections and an automated gap-window rule, was designed for this specific scientific question, and no example of the assembled workflow exists for the agent to imitate. The test is therefore not whether the agent can produce an unseen calculation type. It is whether the agent can assemble familiar components into a new, user-defined research workflow: an executable, dependency-consistent project built to the specification of a scientific question rather than retrieved as a previously seen pattern.

The setup-reconstruction result supplies the evidence: across its twelve cases, LatticeMind regenerated all 23 setup files of this protocol and passed the objective structural and dependency checks, without rerunning the production campaign. The scientific protocol itself was expert-designed; what LatticeMind contributes is the reliable construction and pre-execution validation of the project that carries it out, extending the reliability measured on the benchmark to a workflow the benchmark never contained. This generalization, from reproducing known workflow patterns to constructing a validated project for a previously unseen one, is what separates an agent usable for open-ended research from one confined to documented examples. We turn next to the electronic-structure results.

\subsection{Local interface electronic structure}

\begin{figure*}[tbp]
\centering
\makebox[\textwidth][c]{\includegraphics[width=\FigPhotocatalystPdosTopWidth]{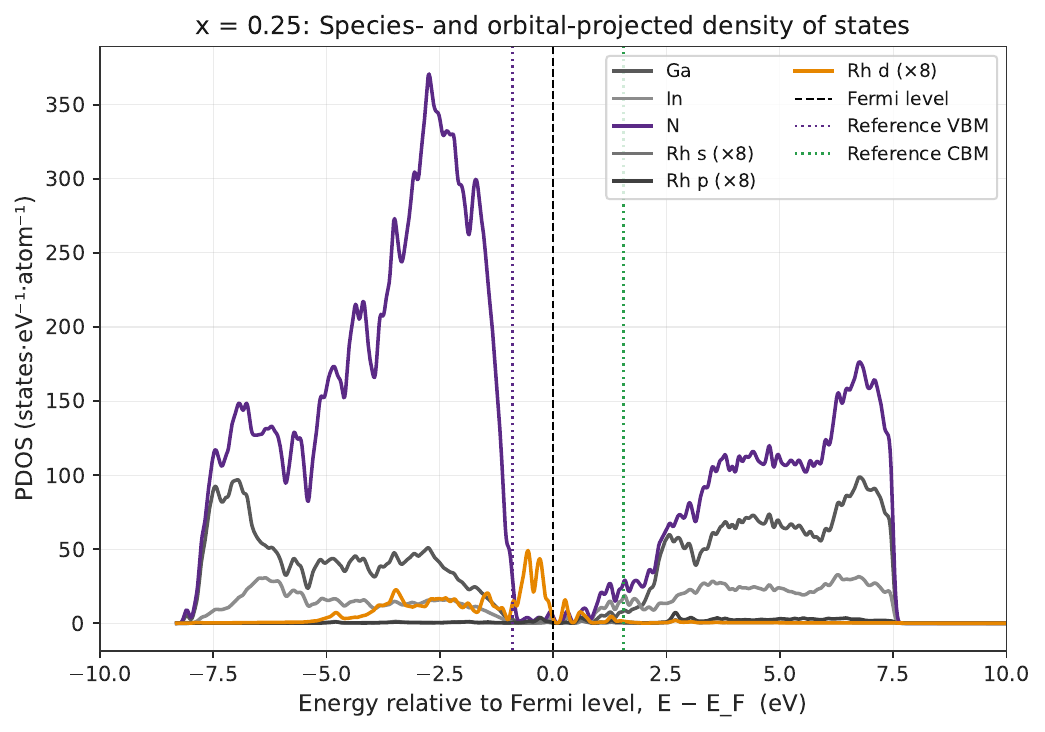}}\\[-0.2ex]
\makebox[\textwidth][c]{\includegraphics[width=\FigPhotocatalystPdosBottomWidth]{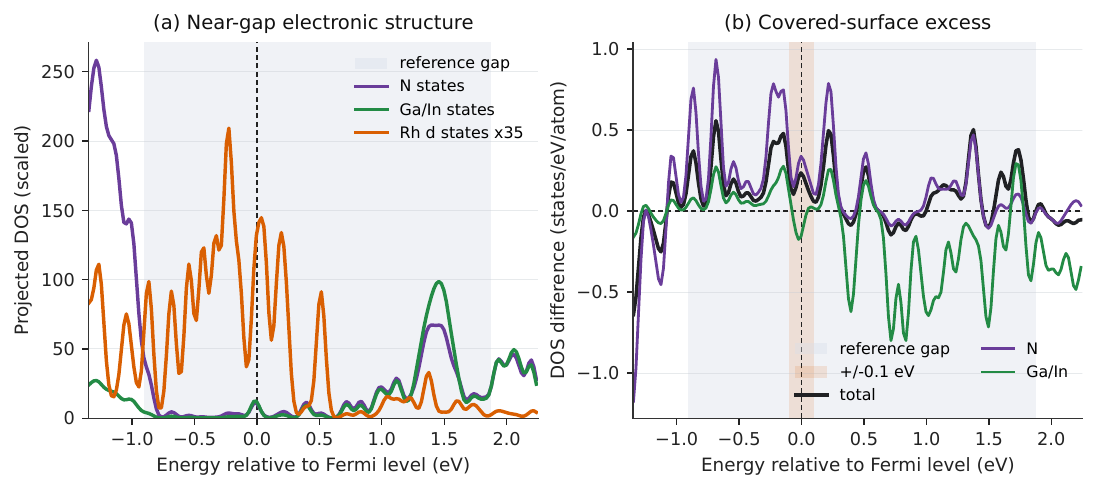}}
\caption{\label{fig:pdos_overview}
Representative projected-DOS analysis for a clean-gap $x=0.25$ SQS configuration. The upper panel shows the broad species- and orbital-projected DOS, and the lower panels show the near-gap electronic structure and covered-minus-uncovered surface PDOS difference. The Fermi level, Region-B reference band-edge markers, reference gap, and $\DNfermi$ window are indicated where relevant. The Rh cluster perturbs the local substrate spectrum beneath the cluster rather than producing a uniform slab-wide shift.}
\end{figure*}

We first examine $x=0.25$, the middle alloy composition, before comparing the alloy trends. Figure~\ref{fig:pdos_overview} shows the broad species- and orbital-projected DOS together with near-gap surface comparisons for a representative clean-gap $x=0.25$ SQS configuration. Near the valence-band edge, N-derived states dominate, whereas Ga/In-derived states dominate the conduction-band edge. The Rh cluster contributes strong near-gap DOS, and its contribution inside the reference gap window is predominantly Rh $d$ in character. This representative picture is supported by the statistics across SQS configurations: all three $x=0.25$ SQS configurations have a clean reference gap, a positive covered-region conduction-onset descriptor, and a Rh $d$ fraction of $0.889\pm0.005$ inside the reference gap window.

The surface PDOS comparison shows larger gap-window DOS in the Rh-covered region than in the uncovered reference region for the representative structure. The difference is concentrated near the gap rather than appearing as a rigid shift of the entire slab. Across the three $x=0.25$ configurations, the covered surface has a positive gap-window excess, $\DNgap=0.228\pm0.115$ states/atom.

\begin{figure}[tbp]
\centering
\includegraphics[width=\FigPhotocatalystDepthWidth]{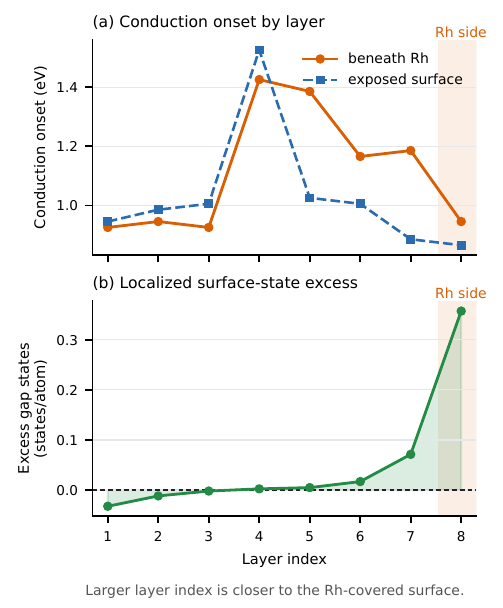}
\caption{\label{fig:depth_metrics}
Depth-resolved descriptors for the representative $x=0.25$ structure. The layer index increases toward the Rh-covered side of the slab. The conduction-onset marker and the gap-state excess both show that the strongest perturbation is localized near the Rh-covered surface.}
\end{figure}

\begin{figure*}[tbp]
\centering
\makebox[\textwidth][c]{\includegraphics[width=\FigPhotocatalystPldosWidth,height=\FigPldosMaxHeight,keepaspectratio]{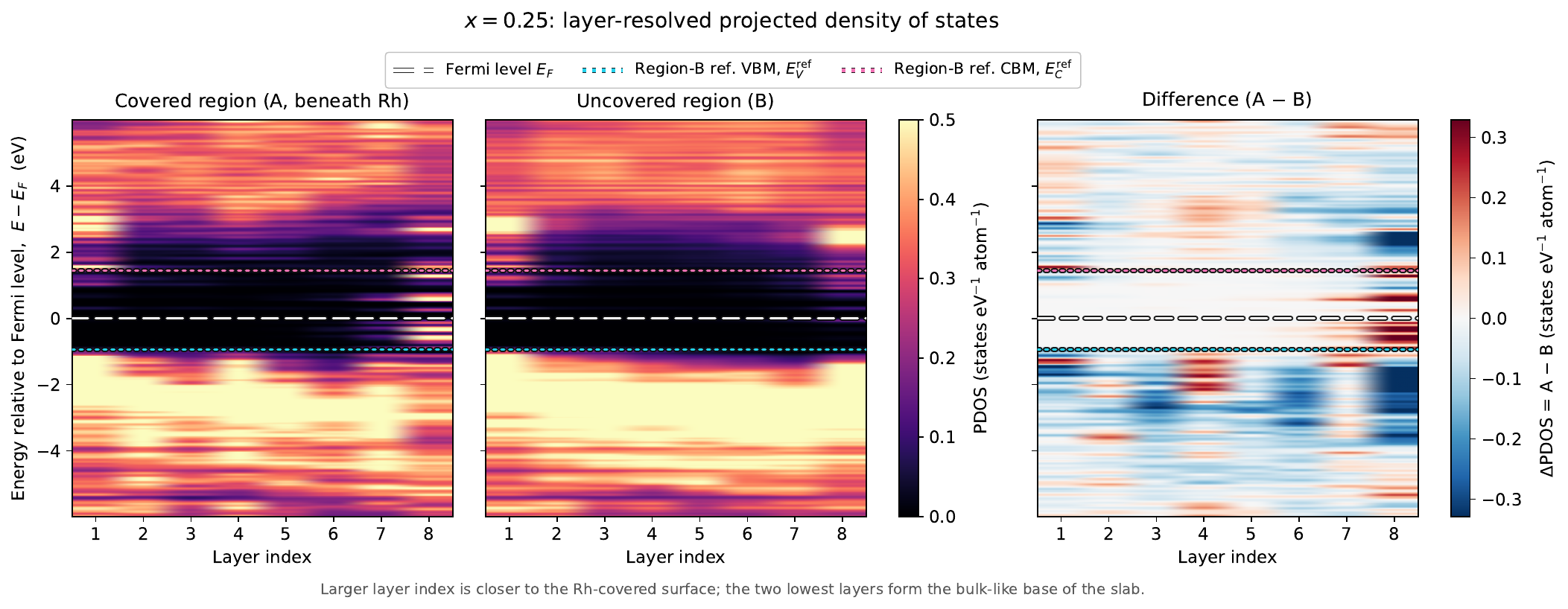}}
\caption{\label{fig:pldos_depth}
Layer-resolved PLDOS for the representative clean-gap $x=0.25$ SQS configuration. The covered region beneath Rh, uncovered region, and covered-minus-uncovered difference show that the surface perturbation is localized near the Rh-covered side of the slab. The layer index increases toward the Rh-covered surface; the two lowest layers ($L=1,2$) are the bulk-like base of the slab and are excluded from the surface-response descriptors. All three panels use linear color scales: the covered (A) and uncovered (B) panels share a common linear PDOS scale, and the covered-minus-uncovered difference uses a linear diverging scale on which the region contrast is read directly. Horizontal markers indicate the Fermi level and the Region-B reference band-edge energies used for the gap-window descriptors.}
\end{figure*}

Figure~\ref{fig:depth_metrics} condenses the same localization into layer-by-layer descriptors. The covered-minus-uncovered difference is largest in the top nitride layers beneath the cluster and decreases toward the slab interior, supporting the laterally heterogeneous interface picture proposed for III-nitride nanocluster photocatalysts~\cite{Yuan2026}. The two deepest layers ($L=1,2$) form the bulk-like base of the slab, farthest from the cluster, and carry little surface or interface physics; the surface-response descriptors are therefore read from the upper layers adjacent to Rh$_6$, and the interior comparison layer $L_{\mathrm{ref}}=5$ and the gap-window search ($L=3$--6) both exclude these deepest layers. The full layer-resolved PLDOS map in Fig.~\ref{fig:pldos_depth} gives the corresponding spectral view.

\subsection{Composition trends}

\begin{figure*}[tbp]
\centering
\includegraphics[width=\FigPhotocatalystTrendsWidth]{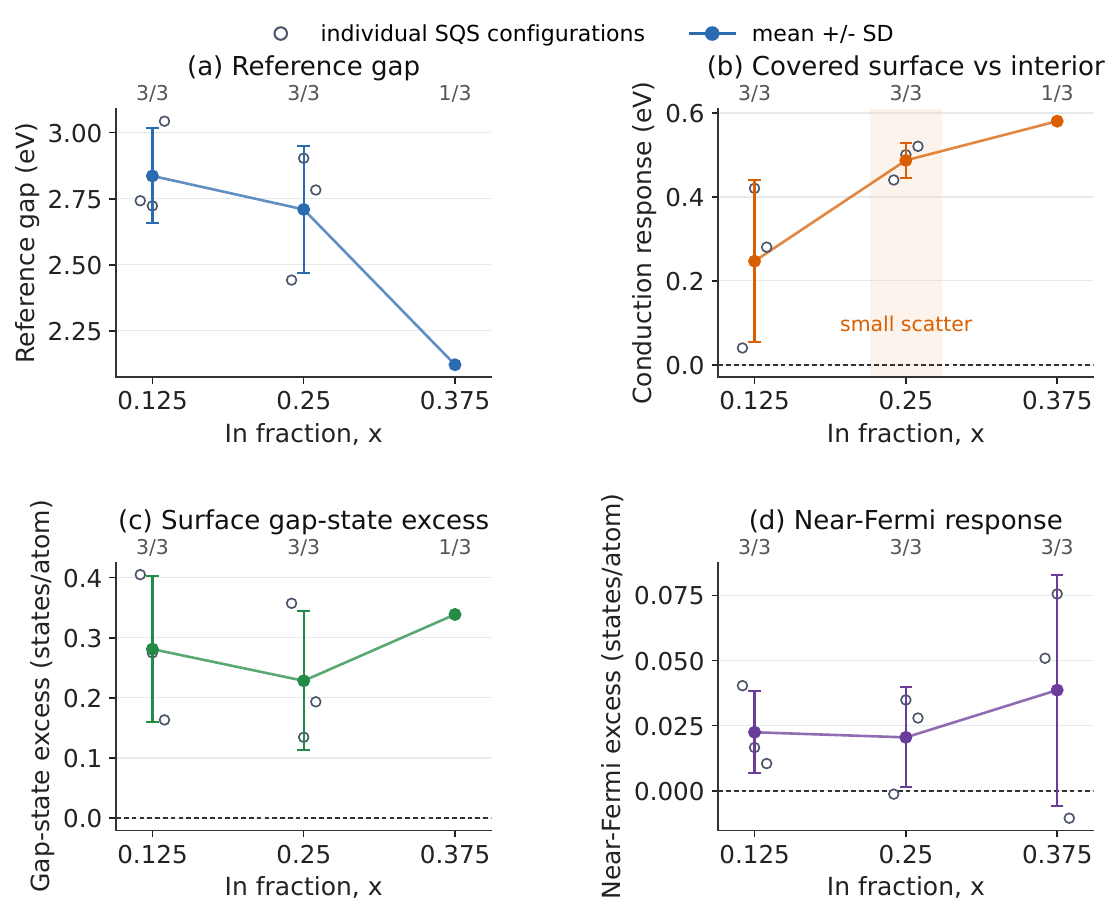}
\caption{\label{fig:trends}
Composition trends over independent SQS configurations for the \RhInGaN\ photocatalyst descriptors. Open symbols are individual SQS configurations, and filled symbols give the mean with one standard deviation where more than one value is available. Gap-window descriptors are plotted only when a clean projected Region-B reference gap is found; labels in panels (a)--(c) give the number of clean-gap configurations out of three. The intermediate alloy, $x=0.25$, gives the most reproducible clean-gap response. The high-In alloy is better described by the near-Fermi metric in panel (d) than by a single clean-gap band-edge shift.}
\end{figure*}

Figure~\ref{fig:trends} summarizes the composition trends from three SQS configurations at each alloy composition, with the full per-composition descriptor values tabulated in the Supplemental Material (Table~S10). Gap-window quantities are reported only for configurations with a clean Region-B reference gap, whereas the near-Fermi descriptor is reported for all configurations. At $x=0.25$, all three SQS configurations have a clean reference gap, with $\Egref=2.71\pm0.24$~eV. The fixed-protocol conduction-onset descriptor is consistently positive, $\DEC=0.487\pm0.042$~eV, meaning that the Region-A conduction onset is lower at the Rh-covered surface than in the Region-A interior. This makes $x=0.25$ the clearest clean-gap case in the sampled alloy set.

At $x=0.125$, the same descriptor remains positive on average, $\DEC=0.247\pm0.192$~eV, but the scatter is large compared with the mean. The low-In structures therefore support the same qualitative covered-region response, but less reproducibly than $x=0.25$. The gap-window surface excess $\DNgap$ is positive in all three $x=0.125$ configurations, with mean $0.281\pm0.121$ states/atom, showing that the covered surface still has more gap-window substrate PDOS than the uncovered surface.

At $x=0.375$, only one of the three SQS configurations has a clean reference gap under the conservative onset rule. The gap-window values listed for this composition, $\Egref=2.12$~eV, $\DEC=0.58$~eV, and $\DNgap=0.339$ states/atom, therefore describe that one gap-defined configuration rather than an average over all three configurations. The two other high-In configurations have projected spectra without a clean Region-B gap by the same rule. For this composition, the near-Fermi covered-minus-uncovered descriptor is more appropriate because it is defined for all three configurations. Its mean is positive, $\DNfermi=0.039\pm0.044$ states/atom, but the scatter includes one slightly negative configuration. Thus the high-In alloy should be described as a near-Fermi surface-response regime rather than as a clean band-edge-shift trend.

Across all gap-defined configurations, the Rh spectral weight inside the reference gap window remains predominantly Rh $d$ in character. The mean Rh $d$ fractions are $0.880\pm0.018$ at $x=0.125$, $0.889\pm0.005$ at $x=0.25$, and $0.873$ for the one gap-defined $x=0.375$ configuration. This supports the assignment of the near-gap covered-surface response to Rh-derived states, while the spread in $\DEC$, $\DNgap$, and $\DNfermi$ shows that the substrate response depends on the local alloy configuration near the Rh footprint.

These results give a composition-dependent but not monotonic picture. Rh produces a local near-gap response beneath the cluster for the gap-defined structures. The response is most reproducible at $x=0.25$, more sensitive to alloy configuration at $x=0.125$, and no longer well represented by a single clean-gap band-edge descriptor at $x=0.375$.

\section{Discussion}

On the RESCU benchmark, LatticeMind reached complete reference-passing for every case and backend without case-specific manual correction, a level of checked workflow success that, to our knowledge, is unprecedented. The ablation shows why: the cumulative build-up assigns the largest incremental gain to the workflow contract, while the leave-one-out comparison assigns the largest loss to removal of the pre-execution validation gate. The reactive trial-and-repair baseline, which has no pre-execution specification to protect the multistage structure and file handoffs, is both less reliable and more expensive per usable result.

These mechanisms also clarify the division of labor inside the architecture. Workflow correctness is enforced by the workflow contract and the deterministic validators, which catch schema, file-dependency, and stage-handoff errors before DFT execution. Physics-level parameter selection and its documented justification are handled by the agents, while physical validity still requires solver-based convergence testing, assessment of model adequacy, and expert review. The researcher agent grounds each project in solver documentation and reference material; the parameterizer agent selects the physics-level settings, such as the $k$-point sampling and the exchange-correlation functional, and records the basis for each choice; and the solver-based convergence testing described in the Introduction remains part of the workflow. The physical reasoning behind a calculation is therefore built into the system and auditable. This lowers the expert effort needed to find appropriate parameters and gives the user documented grounds for evaluating the settings.

The gain quantified in this paper comes from the first of these layers. For large materials campaigns, catching workflow errors before execution changes the failure timing: an invalid generated project is stopped or repaired during the pre-execution checks instead of becoming a solver run that consumes queue time, allocation time, and expert debugging effort.

The \RhInGaN\ case study extends this evidence from benchmark prompts to a real research workflow. Its role is not to serve as an autonomous-discovery benchmark, but to test whether validation-gated construction generalizes to a multistage, composition-dependent protocol that no benchmark case anticipates. The setup-reconstruction campaign regenerated every expected RESCU input file for the documented photocatalyst protocol, and the production calculations preserved the main physical picture: a local Rh-induced, Rh-$d$-dominated near-gap response at the covered surface, strongest and most reproducible in the intermediate alloy. These results are consistent with a laterally heterogeneous interface in which the Rh cluster defines an electronically distinct region beneath itself, while local alloy configuration changes the magnitude and band-edge character of the response.

The practical value of this workflow automation is that it reduces the manual setup and validation burden for families of related DFT calculations. A single DFT setup can be prepared by an expert, but systematic exploration requires repeated changes in composition, SQS configuration, surface region, projection channel, and post-processing window while preserving the same file dependencies and validation records. In the photocatalyst application, LatticeMind-assisted setup lowered the overhead of varying these quantities under one documented RESCU protocol, so the comparisons are controlled by material and configuration changes rather than by uncontrolled changes in workflow definition.

The VASP transfer test clarifies what is solver-specific and what is general. The allowed input fields, documentation source, and helper scripts are adapted to each package, while the validation-gated workflow pattern itself transfers without change.

A natural question is how LatticeMind relates to established workflow managers such as AiiDA~\cite{Pizzi2016} and the atomate/FireWorks ecosystem and its successor atomate2~\cite{Mathew2017,Jain2015}. These are mature and capable systems, but they occupy a different layer of the problem. They are deterministic execution engines: an expert encodes the workflow in advance as code, a WorkChain or a Maker that fixes the stages, settings, dependencies, and error-handling logic, and the framework then runs that predefined graph reliably and records its provenance. The scientific judgment, what to compute, which structure and parameters are appropriate, and how to read the result, is formed in advance by the developers and the user who write that code; the framework executes those judgments but does not form them anew for a new request.

LatticeMind places that reasoning inside the loop. From a natural-language goal alone, its agents interpret the scientific intent, retrieve the relevant solver documentation and reference structures (for example from the Materials Project~\cite{Jain2013MaterialsProject}), decompose the problem into a staged workflow, select the physics-level settings and record the basis for each choice, construct the atomic geometry and, for complex or unsourced cases, vet it through an adversarial reviewer panel, and assemble the dependency-linked project, before the deterministic validators screen it and a report ties the result to its theoretical basis and citations. The workflow-management operations it shares with these frameworks, dependency tracking and saved-density reuse, are a small and downstream part of this process. LatticeMind is therefore not a more elaborate input checker but a computational collaborator that reasons about the calculation, exposing its assumptions and its confidence and deferring to the user where warranted, while the mature managers remain the natural substrate on which to execute and archive the projects it produces. It is this capacity to reason about a calculation, rather than to execute a pre-encoded recipe, that lets it construct a correct workflow for a problem, such as the photocatalyst protocol above, that no one had encoded in advance.

The LatticeMind reliability metrics have a narrow statistical scope. Each main-benchmark prompt was run once, so any benchmark intervals reported in the Supplemental Material (Sec.~S8) describe prompt-level variation across cases rather than repeat variability for the same prompt; the ablation instead uses a 23-case subset with three repeats per case. The direct-LLM and trial-and-repair variants are matched 23-case ablation arms rather than full 44-case baselines. The setup-reconstruction benchmark tests whether LatticeMind can regenerate the input files and dependencies for the documented photocatalyst workflow, not autonomous reproduction of the high-performance-computing campaign behind the production calculations.

The photocatalyst conclusions have a separate physical-model scope. They are PDOS-derived PBE slab estimates on one Rh$_6$ structural motif and one electronic-structure protocol. The alloyed compositions include three SQS configurations each, which is enough to expose local-disorder sensitivity but not enough to exhaustively sample the alloy ensemble. The gap-window descriptors also require a clean projected Region-B gap and are not assigned when that criterion fails. A fully converged predictive model of \RhInGaN\ photocatalysis would additionally require spin-polarized interface checks, explicit solvent and finite-temperature effects, reaction barriers and excited-state dynamics, more Rh$_6$ adsorption motifs, production-level convergence tests, and more extensive sampling of alloy configurations, all of which are listed in the Supplemental Material (Sec.~S10\,E). Within this scope, the qualitative finding of local lateral heterogeneity, the robust intermediate-alloy response, and the architectural conclusions from the ablation remain intact.

\section{Conclusion}

LatticeMind shows that reliable DFT agents should be built around checkable project state: workflow contracts define what must be produced before execution, deterministic validators decide whether the generated project may run, and bounded repair with traceability keeps failures local and auditable. Across the RESCU benchmark, three non-flagship LLM backends each reached this target on every case; the ablation located the gain in the workflow contract and the pre-execution validation gate rather than in model scale; and the same pattern, adapted only in its solver-specific configuration, transferred to VASP, and can be transferred to other DFT packages. 

The backend comparison further suggests that careful workflow architecture can reduce reliance on model scale. This does not replace the need for direct flagship-model comparisons, but it shows that the architecture, rather than raw LLM size, makes smaller-scale non-flagship LLM backends effective. This point is practically important for scientific computing, where cost, latency, data residency, or compute isolation limits the use of large hosted LLM models.

More broadly, computational materials science is shifting from a human-driven loop toward one in which agents translate scientific intent into workflows, execute them, and return results for human interpretation. In that loop the main bottleneck is increasingly not solving the underlying electronic-structure equations but constructing reliable, reproducible computational pipelines at scale. LatticeMind targets exactly this bottleneck, treating scientific reliability as an engineering property that can be specified, checked, and audited. We expect this approach to grow in importance as autonomous and semi-autonomous scientific computing expands.

The \RhInGaN\ application extends this result from benchmark tasks to open-ended use: starting from a natural-language specification, LatticeMind constructed and validated the production-scale multistage project for a research protocol it had never encountered.

The broader lesson is that, for expensive scientific computing, the solver should not be the first validator for errors that are already detectable from the project state. We expect the same design principle to transfer to 
other scientific simulation domains.

\section*{Acknowledgments}
We gratefully acknowledge financial support from the NSERC of Canada and the FRQNT of Quebec (H.G.). The authors thank the Digital Research Alliance of Canada for substantial
computational allocations that made this work possible.
\section*{Data availability}
The benchmark data supporting the reported LatticeMind reliability metrics are described in the Supplemental Material and are publicly available in the LatticeMind benchmark data repository~\cite{Vegh2026LatticeMindData}. Additional materials are available from the corresponding author upon reasonable request.

\vspace{-0.6em}
\makeatletter
\renewcommand{\bibfont}{\fontsize{8}{8.2}\selectfont}
\makeatother
\bibliography{latticemind_prx}

\begin{thebibliography}{39}%
\makeatletter
\providecommand \@ifxundefined [1]{%
 \@ifx{#1\undefined}
}%
\providecommand \@ifnum [1]{%
 \ifnum #1\expandafter \@firstoftwo
 \else \expandafter \@secondoftwo
 \fi
}%
\providecommand \@ifx [1]{%
 \ifx #1\expandafter \@firstoftwo
 \else \expandafter \@secondoftwo
 \fi
}%
\providecommand \natexlab [1]{#1}%
\providecommand \enquote  [1]{``#1''}%
\providecommand \bibnamefont  [1]{#1}%
\providecommand \bibfnamefont [1]{#1}%
\providecommand \citenamefont [1]{#1}%
\providecommand \href@noop [0]{\@secondoftwo}%
\providecommand \href [0]{\begingroup \@sanitize@url \@href}%
\providecommand \@href[1]{\@@startlink{#1}\@@href}%
\providecommand \@@href[1]{\endgroup#1\@@endlink}%
\providecommand \@sanitize@url [0]{\catcode `\\12\catcode `\$12\catcode
  `\&12\catcode `\#12\catcode `\^12\catcode `\_12\catcode `\%12\relax}%
\providecommand \@@startlink[1]{}%
\providecommand \@@endlink[0]{}%
\providecommand \url  [0]{\begingroup\@sanitize@url \@url }%
\providecommand \@url [1]{\endgroup\@href {#1}{\urlprefix }}%
\providecommand \urlprefix  [0]{URL }%
\providecommand \Eprint [0]{\href }%
\providecommand \doibase [0]{https://doi.org/}%
\providecommand \selectlanguage [0]{\@gobble}%
\providecommand \bibinfo  [0]{\@secondoftwo}%
\providecommand \bibfield  [0]{\@secondoftwo}%
\providecommand \translation [1]{[#1]}%
\providecommand \BibitemOpen [0]{}%
\providecommand \bibitemStop [0]{}%
\providecommand \bibitemNoStop [0]{.\EOS\space}%
\providecommand \EOS [0]{\spacefactor3000\relax}%
\providecommand \BibitemShut  [1]{\csname bibitem#1\endcsname}%
\let\auto@bib@innerbib\@empty
\bibitem [{\citenamefont {Hohenberg}\ and\ \citenamefont
  {Kohn}(1964)}]{Hohenberg1964}%
  \BibitemOpen
  \bibfield  {author} {\bibinfo {author} {\bibfnamefont {P.}~\bibnamefont
  {Hohenberg}}\ and\ \bibinfo {author} {\bibfnamefont {W.}~\bibnamefont
  {Kohn}},\ }\bibfield  {title} {\bibinfo {title} {Inhomogeneous electron
  gas},\ }\href {https://doi.org/10.1103/PhysRev.136.B864} {\bibfield
  {journal} {\bibinfo  {journal} {Phys. Rev.}\ }\textbf {\bibinfo {volume}
  {136}},\ \bibinfo {pages} {B864} (\bibinfo {year} {1964})}\BibitemShut
  {NoStop}%
\bibitem [{\citenamefont {Kohn}\ and\ \citenamefont {Sham}(1965)}]{Kohn1965}%
  \BibitemOpen
  \bibfield  {author} {\bibinfo {author} {\bibfnamefont {W.}~\bibnamefont
  {Kohn}}\ and\ \bibinfo {author} {\bibfnamefont {L.~J.}\ \bibnamefont
  {Sham}},\ }\bibfield  {title} {\bibinfo {title} {Self-consistent equations
  including exchange and correlation effects},\ }\href
  {https://doi.org/10.1103/PhysRev.140.A1133} {\bibfield  {journal} {\bibinfo
  {journal} {Phys. Rev.}\ }\textbf {\bibinfo {volume} {140}},\ \bibinfo {pages}
  {A1133} (\bibinfo {year} {1965})}\BibitemShut {NoStop}%
\bibitem [{\citenamefont {Jones}(2015)}]{Jones2015DFTReview}%
  \BibitemOpen
  \bibfield  {author} {\bibinfo {author} {\bibfnamefont {R.~O.}\ \bibnamefont
  {Jones}},\ }\bibfield  {title} {\bibinfo {title} {Density functional theory:
  Its origins, rise to prominence, and future},\ }\href
  {https://doi.org/10.1103/RevModPhys.87.897} {\bibfield  {journal} {\bibinfo
  {journal} {Reviews of Modern Physics}\ }\textbf {\bibinfo {volume} {87}},\
  \bibinfo {pages} {897} (\bibinfo {year} {2015})}\BibitemShut {NoStop}%
\bibitem [{\citenamefont {{National Energy Research Scientific Computing
  Center}}(2024)}]{NERSC2023AnnualReport}%
  \BibitemOpen
  \bibfield  {author} {\bibinfo {author} {\bibnamefont {{National Energy
  Research Scientific Computing Center}}},\ }\href
  {https://www.nersc.gov/assets/Uploads/Elements/FileList/Annual-Reports/2023-NERSC-Annual-Report.pdf}
  {\bibinfo {title} {{NERSC} annual report 2023}} (\bibinfo {year}
  {2024})\BibitemShut {NoStop}%
\bibitem [{\citenamefont {Zhao}\ \emph {et~al.}(2023)\citenamefont {Zhao},
  \citenamefont {Austin}, \citenamefont {Maintz},\ and\ \citenamefont
  {Marsman}}]{Zhao2023VASPPerlmutter}%
  \BibitemOpen
  \bibfield  {author} {\bibinfo {author} {\bibfnamefont {Z.}~\bibnamefont
  {Zhao}}, \bibinfo {author} {\bibfnamefont {B.}~\bibnamefont {Austin}},
  \bibinfo {author} {\bibfnamefont {S.}~\bibnamefont {Maintz}},\ and\ \bibinfo
  {author} {\bibfnamefont {M.}~\bibnamefont {Marsman}},\ }\bibfield  {title}
  {\bibinfo {title} {{VASP} performance on {HPE} {Cray} {EX} based on {NVIDIA}
  {A100} {GPUs} and {AMD} {Milan} {CPUs}},\ }in\ \href
  {https://cug.org/proceedings/cug2023_proceedings/includes/files/pap130s2-file1.pdf}
  {\emph {\bibinfo {booktitle} {Proceedings of the Cray User Group Meeting}}}\
  (\bibinfo {year} {2023})\BibitemShut {NoStop}%
\bibitem [{\citenamefont {Kirklin}\ \emph {et~al.}(2015)\citenamefont
  {Kirklin}, \citenamefont {Saal}, \citenamefont {Meredig}, \citenamefont
  {Thompson}, \citenamefont {Doak}, \citenamefont {Aykol}, \citenamefont
  {R{\"u}hl},\ and\ \citenamefont {Wolverton}}]{Kirklin2015OQMD}%
  \BibitemOpen
  \bibfield  {author} {\bibinfo {author} {\bibfnamefont {S.}~\bibnamefont
  {Kirklin}}, \bibinfo {author} {\bibfnamefont {J.~E.}\ \bibnamefont {Saal}},
  \bibinfo {author} {\bibfnamefont {B.}~\bibnamefont {Meredig}}, \bibinfo
  {author} {\bibfnamefont {A.}~\bibnamefont {Thompson}}, \bibinfo {author}
  {\bibfnamefont {J.~W.}\ \bibnamefont {Doak}}, \bibinfo {author}
  {\bibfnamefont {M.}~\bibnamefont {Aykol}}, \bibinfo {author} {\bibfnamefont
  {S.}~\bibnamefont {R{\"u}hl}},\ and\ \bibinfo {author} {\bibfnamefont
  {C.}~\bibnamefont {Wolverton}},\ }\bibfield  {title} {\bibinfo {title} {The
  open quantum materials database ({OQMD}): Assessing the accuracy of {DFT}
  formation energies},\ }\href {https://doi.org/10.1038/npjcompumats.2015.10}
  {\bibfield  {journal} {\bibinfo  {journal} {npj Comput. Mater.}\ }\textbf
  {\bibinfo {volume} {1}},\ \bibinfo {pages} {15010} (\bibinfo {year}
  {2015})}\BibitemShut {NoStop}%
\bibitem [{\citenamefont {Schmidt}\ \emph {et~al.}(2022)\citenamefont
  {Schmidt}, \citenamefont {Wang}, \citenamefont {Cerqueira}, \citenamefont
  {Botti},\ and\ \citenamefont {Marques}}]{Schmidt2022PBESolSCAN}%
  \BibitemOpen
  \bibfield  {author} {\bibinfo {author} {\bibfnamefont {J.}~\bibnamefont
  {Schmidt}}, \bibinfo {author} {\bibfnamefont {H.-C.}\ \bibnamefont {Wang}},
  \bibinfo {author} {\bibfnamefont {T.~F.~T.}\ \bibnamefont {Cerqueira}},
  \bibinfo {author} {\bibfnamefont {S.}~\bibnamefont {Botti}},\ and\ \bibinfo
  {author} {\bibfnamefont {M.~A.~L.}\ \bibnamefont {Marques}},\ }\bibfield
  {title} {\bibinfo {title} {A dataset of 175k stable and metastable materials
  calculated with the {PBEsol} and {SCAN} functionals},\ }\href
  {https://doi.org/10.1038/s41597-022-01177-w} {\bibfield  {journal} {\bibinfo
  {journal} {Sci. Data}\ }\textbf {\bibinfo {volume} {9}},\ \bibinfo {pages}
  {64} (\bibinfo {year} {2022})}\BibitemShut {NoStop}%
\bibitem [{\citenamefont {Deng}\ \emph {et~al.}(2023)\citenamefont {Deng},
  \citenamefont {Zhong}, \citenamefont {Jun}, \citenamefont {Riebesell},
  \citenamefont {Han}, \citenamefont {Bartel},\ and\ \citenamefont
  {Ceder}}]{Deng2023CHGNet}%
  \BibitemOpen
  \bibfield  {author} {\bibinfo {author} {\bibfnamefont {B.}~\bibnamefont
  {Deng}}, \bibinfo {author} {\bibfnamefont {P.}~\bibnamefont {Zhong}},
  \bibinfo {author} {\bibfnamefont {K.}~\bibnamefont {Jun}}, \bibinfo {author}
  {\bibfnamefont {J.}~\bibnamefont {Riebesell}}, \bibinfo {author}
  {\bibfnamefont {K.}~\bibnamefont {Han}}, \bibinfo {author} {\bibfnamefont
  {C.~J.}\ \bibnamefont {Bartel}},\ and\ \bibinfo {author} {\bibfnamefont
  {G.}~\bibnamefont {Ceder}},\ }\bibfield  {title} {\bibinfo {title} {{CHGNet}
  as a pretrained universal neural network potential for charge-informed
  atomistic modelling},\ }\href {https://doi.org/10.1038/s42256-023-00716-3}
  {\bibfield  {journal} {\bibinfo  {journal} {Nature Machine Intelligence}\
  }\textbf {\bibinfo {volume} {5}},\ \bibinfo {pages} {1031} (\bibinfo {year}
  {2023})}\BibitemShut {NoStop}%
\bibitem [{\citenamefont {Boiko}\ \emph {et~al.}(2023)\citenamefont {Boiko},
  \citenamefont {MacKnight}, \citenamefont {Kline},\ and\ \citenamefont
  {Gomes}}]{Boiko2023}%
  \BibitemOpen
  \bibfield  {author} {\bibinfo {author} {\bibfnamefont {D.~A.}\ \bibnamefont
  {Boiko}}, \bibinfo {author} {\bibfnamefont {R.}~\bibnamefont {MacKnight}},
  \bibinfo {author} {\bibfnamefont {B.}~\bibnamefont {Kline}},\ and\ \bibinfo
  {author} {\bibfnamefont {G.}~\bibnamefont {Gomes}},\ }\bibfield  {title}
  {\bibinfo {title} {Autonomous chemical research with large language models},\
  }\href {https://doi.org/10.1038/s41586-023-06792-0} {\bibfield  {journal}
  {\bibinfo  {journal} {Nature}\ }\textbf {\bibinfo {volume} {624}},\ \bibinfo
  {pages} {570} (\bibinfo {year} {2023})}\BibitemShut {NoStop}%
\bibitem [{\citenamefont {Bran}\ \emph {et~al.}(2024)\citenamefont {Bran},
  \citenamefont {Cox}, \citenamefont {Schilter}, \citenamefont {Baldassari},
  \citenamefont {White},\ and\ \citenamefont {Schwaller}}]{Bran2024}%
  \BibitemOpen
  \bibfield  {author} {\bibinfo {author} {\bibfnamefont {A.~M.}\ \bibnamefont
  {Bran}}, \bibinfo {author} {\bibfnamefont {S.}~\bibnamefont {Cox}}, \bibinfo
  {author} {\bibfnamefont {O.}~\bibnamefont {Schilter}}, \bibinfo {author}
  {\bibfnamefont {C.}~\bibnamefont {Baldassari}}, \bibinfo {author}
  {\bibfnamefont {A.~D.}\ \bibnamefont {White}},\ and\ \bibinfo {author}
  {\bibfnamefont {P.}~\bibnamefont {Schwaller}},\ }\bibfield  {title} {\bibinfo
  {title} {Augmenting large language models with chemistry tools},\ }\href
  {https://doi.org/10.1038/s42256-024-00832-8} {\bibfield  {journal} {\bibinfo
  {journal} {Nat. Mach. Intell.}\ }\textbf {\bibinfo {volume} {6}},\ \bibinfo
  {pages} {525} (\bibinfo {year} {2024})}\BibitemShut {NoStop}%
\bibitem [{\citenamefont {Ansari}\ and\ \citenamefont
  {Moosavi}(2024)}]{Ansari2024}%
  \BibitemOpen
  \bibfield  {author} {\bibinfo {author} {\bibfnamefont {M.}~\bibnamefont
  {Ansari}}\ and\ \bibinfo {author} {\bibfnamefont {S.~M.}\ \bibnamefont
  {Moosavi}},\ }\bibfield  {title} {\bibinfo {title} {Agent-based learning of
  materials datasets from the scientific literature},\ }\href
  {https://doi.org/10.1039/D4DD00252K} {\bibfield  {journal} {\bibinfo
  {journal} {Digital Discovery}\ }\textbf {\bibinfo {volume} {3}},\ \bibinfo
  {pages} {2607} (\bibinfo {year} {2024})}\BibitemShut {NoStop}%
\bibitem [{\citenamefont {Mirza}\ \emph {et~al.}(2025)\citenamefont {Mirza},
  \citenamefont {Alampara}, \citenamefont {Kunchapu}, \citenamefont
  {R{\'i}os-Garc{\'i}a} \emph {et~al.}}]{Mirza2025}%
  \BibitemOpen
  \bibfield  {author} {\bibinfo {author} {\bibfnamefont {A.}~\bibnamefont
  {Mirza}}, \bibinfo {author} {\bibfnamefont {N.}~\bibnamefont {Alampara}},
  \bibinfo {author} {\bibfnamefont {S.}~\bibnamefont {Kunchapu}}, \bibinfo
  {author} {\bibfnamefont {M.}~\bibnamefont {R{\'i}os-Garc{\'i}a}}, \emph
  {et~al.},\ }\bibfield  {title} {\bibinfo {title} {A framework for evaluating
  the chemical knowledge and reasoning abilities of large language models
  against the expertise of chemists},\ }\href
  {https://doi.org/10.1038/s41557-025-01815-x} {\bibfield  {journal} {\bibinfo
  {journal} {Nat. Chem.}\ }\textbf {\bibinfo {volume} {17}},\ \bibinfo {pages}
  {1027} (\bibinfo {year} {2025})}\BibitemShut {NoStop}%
\bibitem [{\citenamefont {Mandal}\ \emph {et~al.}(2025)\citenamefont {Mandal},
  \citenamefont {Soni}, \citenamefont {Zaki}, \citenamefont {Smedskjaer},
  \citenamefont {Wondraczek}, \citenamefont {Wondraczek}, \citenamefont
  {Gosvami},\ and\ \citenamefont {Krishnan}}]{Mandal2025}%
  \BibitemOpen
  \bibfield  {author} {\bibinfo {author} {\bibfnamefont {I.}~\bibnamefont
  {Mandal}}, \bibinfo {author} {\bibfnamefont {J.}~\bibnamefont {Soni}},
  \bibinfo {author} {\bibfnamefont {M.}~\bibnamefont {Zaki}}, \bibinfo {author}
  {\bibfnamefont {M.~M.}\ \bibnamefont {Smedskjaer}}, \bibinfo {author}
  {\bibfnamefont {K.}~\bibnamefont {Wondraczek}}, \bibinfo {author}
  {\bibfnamefont {L.}~\bibnamefont {Wondraczek}}, \bibinfo {author}
  {\bibfnamefont {N.~N.}\ \bibnamefont {Gosvami}},\ and\ \bibinfo {author}
  {\bibfnamefont {N.~M.~A.}\ \bibnamefont {Krishnan}},\ }\bibfield  {title}
  {\bibinfo {title} {Evaluating large language model agents for automation of
  atomic force microscopy},\ }\href
  {https://doi.org/10.1038/s41467-025-64105-7} {\bibfield  {journal} {\bibinfo
  {journal} {Nat. Commun.}\ }\textbf {\bibinfo {volume} {16}},\ \bibinfo
  {pages} {9104} (\bibinfo {year} {2025})}\BibitemShut {NoStop}%
\bibitem [{\citenamefont {Zou}\ \emph {et~al.}(2025)\citenamefont {Zou},
  \citenamefont {Cheng}, \citenamefont {Aldossary}, \citenamefont {Bai},
  \citenamefont {Leong}, \citenamefont {Campos-Gonzalez-Angulo}, \citenamefont
  {Choi}, \citenamefont {Ser}, \citenamefont {Tom}, \citenamefont {Wang},
  \citenamefont {Zhang}, \citenamefont {Yakavets}, \citenamefont {Hao},
  \citenamefont {Crebolder}, \citenamefont {Bernales},\ and\ \citenamefont
  {Aspuru-Guzik}}]{Zou2025}%
  \BibitemOpen
  \bibfield  {author} {\bibinfo {author} {\bibfnamefont {Y.}~\bibnamefont
  {Zou}}, \bibinfo {author} {\bibfnamefont {A.~H.}\ \bibnamefont {Cheng}},
  \bibinfo {author} {\bibfnamefont {A.}~\bibnamefont {Aldossary}}, \bibinfo
  {author} {\bibfnamefont {J.}~\bibnamefont {Bai}}, \bibinfo {author}
  {\bibfnamefont {S.~X.}\ \bibnamefont {Leong}}, \bibinfo {author}
  {\bibfnamefont {J.~A.}\ \bibnamefont {Campos-Gonzalez-Angulo}}, \bibinfo
  {author} {\bibfnamefont {C.}~\bibnamefont {Choi}}, \bibinfo {author}
  {\bibfnamefont {C.~T.}\ \bibnamefont {Ser}}, \bibinfo {author} {\bibfnamefont
  {G.}~\bibnamefont {Tom}}, \bibinfo {author} {\bibfnamefont {A.}~\bibnamefont
  {Wang}}, \bibinfo {author} {\bibfnamefont {Z.}~\bibnamefont {Zhang}},
  \bibinfo {author} {\bibfnamefont {I.}~\bibnamefont {Yakavets}}, \bibinfo
  {author} {\bibfnamefont {H.}~\bibnamefont {Hao}}, \bibinfo {author}
  {\bibfnamefont {C.}~\bibnamefont {Crebolder}}, \bibinfo {author}
  {\bibfnamefont {V.}~\bibnamefont {Bernales}},\ and\ \bibinfo {author}
  {\bibfnamefont {A.}~\bibnamefont {Aspuru-Guzik}},\ }\bibfield  {title}
  {\bibinfo {title} {El agente: An autonomous agent for quantum chemistry},\
  }\href {https://doi.org/10.1016/j.matt.2025.102263} {\bibfield  {journal}
  {\bibinfo  {journal} {Matter}\ }\textbf {\bibinfo {volume} {8}},\ \bibinfo
  {pages} {102263} (\bibinfo {year} {2025})}\BibitemShut {NoStop}%
\bibitem [{\citenamefont {Jain}\ \emph {et~al.}(2015)\citenamefont {Jain},
  \citenamefont {Ong}, \citenamefont {Chen}, \citenamefont {Medasani},
  \citenamefont {Qu}, \citenamefont {Kocher}, \citenamefont {Brafman},
  \citenamefont {Petretto}, \citenamefont {Rignanese}, \citenamefont {Hautier},
  \citenamefont {Gunter},\ and\ \citenamefont {Persson}}]{Jain2015}%
  \BibitemOpen
  \bibfield  {author} {\bibinfo {author} {\bibfnamefont {A.}~\bibnamefont
  {Jain}}, \bibinfo {author} {\bibfnamefont {S.~P.}\ \bibnamefont {Ong}},
  \bibinfo {author} {\bibfnamefont {W.}~\bibnamefont {Chen}}, \bibinfo {author}
  {\bibfnamefont {B.}~\bibnamefont {Medasani}}, \bibinfo {author}
  {\bibfnamefont {X.}~\bibnamefont {Qu}}, \bibinfo {author} {\bibfnamefont
  {M.}~\bibnamefont {Kocher}}, \bibinfo {author} {\bibfnamefont
  {M.}~\bibnamefont {Brafman}}, \bibinfo {author} {\bibfnamefont
  {G.}~\bibnamefont {Petretto}}, \bibinfo {author} {\bibfnamefont {G.-M.}\
  \bibnamefont {Rignanese}}, \bibinfo {author} {\bibfnamefont {G.}~\bibnamefont
  {Hautier}}, \bibinfo {author} {\bibfnamefont {D.}~\bibnamefont {Gunter}},\
  and\ \bibinfo {author} {\bibfnamefont {K.~A.}\ \bibnamefont {Persson}},\
  }\bibfield  {title} {\bibinfo {title} {Fireworks: A dynamic workflow system
  designed for high-throughput applications},\ }\href
  {https://doi.org/10.1002/cpe.3505} {\bibfield  {journal} {\bibinfo  {journal}
  {Concurrency Computat.: Pract. Exper.}\ }\textbf {\bibinfo {volume} {27}},\
  \bibinfo {pages} {5037} (\bibinfo {year} {2015})}\BibitemShut {NoStop}%
\bibitem [{\citenamefont {Mathew}\ \emph {et~al.}(2017)\citenamefont {Mathew},
  \citenamefont {Montoya}, \citenamefont {Faghaninia}, \citenamefont
  {Dwarakanath}, \citenamefont {Aykol}, \citenamefont {Tang}, \citenamefont
  {Chu}, \citenamefont {Smidt}, \citenamefont {Bocklund}, \citenamefont
  {Horton}, \citenamefont {Dagdelen}, \citenamefont {Wood}, \citenamefont
  {Liu}, \citenamefont {Neaton}, \citenamefont {Ong}, \citenamefont {Persson},\
  and\ \citenamefont {Jain}}]{Mathew2017}%
  \BibitemOpen
  \bibfield  {author} {\bibinfo {author} {\bibfnamefont {K.}~\bibnamefont
  {Mathew}}, \bibinfo {author} {\bibfnamefont {J.~H.}\ \bibnamefont {Montoya}},
  \bibinfo {author} {\bibfnamefont {A.}~\bibnamefont {Faghaninia}}, \bibinfo
  {author} {\bibfnamefont {S.}~\bibnamefont {Dwarakanath}}, \bibinfo {author}
  {\bibfnamefont {M.}~\bibnamefont {Aykol}}, \bibinfo {author} {\bibfnamefont
  {H.}~\bibnamefont {Tang}}, \bibinfo {author} {\bibfnamefont {I.-H.}\
  \bibnamefont {Chu}}, \bibinfo {author} {\bibfnamefont {T.}~\bibnamefont
  {Smidt}}, \bibinfo {author} {\bibfnamefont {B.}~\bibnamefont {Bocklund}},
  \bibinfo {author} {\bibfnamefont {M.}~\bibnamefont {Horton}}, \bibinfo
  {author} {\bibfnamefont {J.}~\bibnamefont {Dagdelen}}, \bibinfo {author}
  {\bibfnamefont {B.}~\bibnamefont {Wood}}, \bibinfo {author} {\bibfnamefont
  {Z.-K.}\ \bibnamefont {Liu}}, \bibinfo {author} {\bibfnamefont
  {J.}~\bibnamefont {Neaton}}, \bibinfo {author} {\bibfnamefont {S.~P.}\
  \bibnamefont {Ong}}, \bibinfo {author} {\bibfnamefont {K.~A.}\ \bibnamefont
  {Persson}},\ and\ \bibinfo {author} {\bibfnamefont {A.}~\bibnamefont
  {Jain}},\ }\bibfield  {title} {\bibinfo {title} {Atomate: A high-level
  interface to generate, execute, and analyze computational materials science
  workflows},\ }\href {https://doi.org/10.1016/j.commatsci.2017.07.030}
  {\bibfield  {journal} {\bibinfo  {journal} {Comput. Mater. Sci.}\ }\textbf
  {\bibinfo {volume} {139}},\ \bibinfo {pages} {140} (\bibinfo {year}
  {2017})}\BibitemShut {NoStop}%
\bibitem [{\citenamefont {Pizzi}\ \emph {et~al.}(2016)\citenamefont {Pizzi},
  \citenamefont {Cepellotti}, \citenamefont {Sabatini}, \citenamefont
  {Marzari},\ and\ \citenamefont {Kozinsky}}]{Pizzi2016}%
  \BibitemOpen
  \bibfield  {author} {\bibinfo {author} {\bibfnamefont {G.}~\bibnamefont
  {Pizzi}}, \bibinfo {author} {\bibfnamefont {A.}~\bibnamefont {Cepellotti}},
  \bibinfo {author} {\bibfnamefont {R.}~\bibnamefont {Sabatini}}, \bibinfo
  {author} {\bibfnamefont {N.}~\bibnamefont {Marzari}},\ and\ \bibinfo {author}
  {\bibfnamefont {B.}~\bibnamefont {Kozinsky}},\ }\bibfield  {title} {\bibinfo
  {title} {{AiiDA}: Automated interactive infrastructure and database for
  computational science},\ }\href
  {https://doi.org/10.1016/j.commatsci.2015.09.013} {\bibfield  {journal}
  {\bibinfo  {journal} {Comput. Mater. Sci.}\ }\textbf {\bibinfo {volume}
  {111}},\ \bibinfo {pages} {218} (\bibinfo {year} {2016})}\BibitemShut
  {NoStop}%
\bibitem [{\citenamefont {Blau}\ \emph {et~al.}(2020)\citenamefont {Blau},
  \citenamefont {Spotte-Smith}, \citenamefont {Wood}, \citenamefont
  {Dwaraknath},\ and\ \citenamefont {Persson}}]{Blau2020DFTErrorCorrection}%
  \BibitemOpen
  \bibfield  {author} {\bibinfo {author} {\bibfnamefont {S.}~\bibnamefont
  {Blau}}, \bibinfo {author} {\bibfnamefont {E.~W.~C.}\ \bibnamefont
  {Spotte-Smith}}, \bibinfo {author} {\bibfnamefont {B.}~\bibnamefont {Wood}},
  \bibinfo {author} {\bibfnamefont {S.}~\bibnamefont {Dwaraknath}},\ and\
  \bibinfo {author} {\bibfnamefont {K.}~\bibnamefont {Persson}},\ }\bibfield
  {title} {\bibinfo {title} {Accurate, automated density functional theory for
  complex molecules using on-the-fly error correction},\ }\bibfield  {journal}
  {\bibinfo  {journal} {ChemRxiv}\ }\href
  {https://doi.org/10.26434/chemrxiv.13076030.v1}
  {10.26434/chemrxiv.13076030.v1} (\bibinfo {year} {2020})\BibitemShut
  {NoStop}%
\bibitem [{\citenamefont {Huber}\ \emph {et~al.}(2026)\citenamefont {Huber},
  \citenamefont {Minotakis}, \citenamefont {Bercx}, \citenamefont {Reents},
  \citenamefont {Eimre}, \citenamefont {Paulish}, \citenamefont {H{\"o}rmann},
  \citenamefont {Uhrin}, \citenamefont {Marzari},\ and\ \citenamefont
  {Pizzi}}]{Huber2026MC3D}%
  \BibitemOpen
  \bibfield  {author} {\bibinfo {author} {\bibfnamefont {S.~P.}\ \bibnamefont
  {Huber}}, \bibinfo {author} {\bibfnamefont {M.}~\bibnamefont {Minotakis}},
  \bibinfo {author} {\bibfnamefont {M.}~\bibnamefont {Bercx}}, \bibinfo
  {author} {\bibfnamefont {T.}~\bibnamefont {Reents}}, \bibinfo {author}
  {\bibfnamefont {K.}~\bibnamefont {Eimre}}, \bibinfo {author} {\bibfnamefont
  {N.}~\bibnamefont {Paulish}}, \bibinfo {author} {\bibfnamefont
  {N.}~\bibnamefont {H{\"o}rmann}}, \bibinfo {author} {\bibfnamefont
  {M.}~\bibnamefont {Uhrin}}, \bibinfo {author} {\bibfnamefont
  {N.}~\bibnamefont {Marzari}},\ and\ \bibinfo {author} {\bibfnamefont
  {G.}~\bibnamefont {Pizzi}},\ }\bibfield  {title} {\bibinfo {title} {{MC3D}:
  The materials cloud computational database of experimentally known
  stoichiometric inorganics},\ }\href {https://doi.org/10.1039/D5DD00415B}
  {\bibfield  {journal} {\bibinfo  {journal} {Digital Discovery}\ }\textbf
  {\bibinfo {volume} {5}},\ \bibinfo {pages} {1114} (\bibinfo {year}
  {2026})}\BibitemShut {NoStop}%
\bibitem [{\citenamefont {Gugler}\ \emph {et~al.}(2020)\citenamefont {Gugler},
  \citenamefont {Janet},\ and\ \citenamefont
  {Kulik}}]{Gugler2020DeNovoInorganic}%
  \BibitemOpen
  \bibfield  {author} {\bibinfo {author} {\bibfnamefont {S.}~\bibnamefont
  {Gugler}}, \bibinfo {author} {\bibfnamefont {J.~P.}\ \bibnamefont {Janet}},\
  and\ \bibinfo {author} {\bibfnamefont {H.~J.}\ \bibnamefont {Kulik}},\
  }\bibfield  {title} {\bibinfo {title} {Enumeration of de novo inorganic
  complexes for chemical discovery and machine learning},\ }\href
  {https://doi.org/10.1039/C9ME00069K} {\bibfield  {journal} {\bibinfo
  {journal} {Mol. Syst. Des. Eng.}\ }\textbf {\bibinfo {volume} {5}},\ \bibinfo
  {pages} {139} (\bibinfo {year} {2020})}\BibitemShut {NoStop}%
\bibitem [{\citenamefont {Yuan}\ \emph {et~al.}(2012)\citenamefont {Yuan},
  \citenamefont {Wu}, \citenamefont {Wang}, \citenamefont {Yang},\ and\
  \citenamefont {Zheng}}]{Yuan2012HPCFailures}%
  \BibitemOpen
  \bibfield  {author} {\bibinfo {author} {\bibfnamefont {Y.}~\bibnamefont
  {Yuan}}, \bibinfo {author} {\bibfnamefont {Y.}~\bibnamefont {Wu}}, \bibinfo
  {author} {\bibfnamefont {Q.}~\bibnamefont {Wang}}, \bibinfo {author}
  {\bibfnamefont {G.}~\bibnamefont {Yang}},\ and\ \bibinfo {author}
  {\bibfnamefont {W.}~\bibnamefont {Zheng}},\ }\bibfield  {title} {\bibinfo
  {title} {Job failures in high performance computing systems: A large-scale
  empirical study},\ }\href {https://doi.org/10.1016/j.camwa.2011.07.040}
  {\bibfield  {journal} {\bibinfo  {journal} {Comput. Math. Appl.}\ }\textbf
  {\bibinfo {volume} {63}},\ \bibinfo {pages} {365} (\bibinfo {year}
  {2012})}\BibitemShut {NoStop}%
\bibitem [{\citenamefont {Hu}\ \emph {et~al.}(2026)\citenamefont {Hu},
  \citenamefont {Talit}, \citenamefont {Wang}, \citenamefont {Ahmad},
  \citenamefont {Lin}, \citenamefont {Kaur}, \citenamefont {Lane},
  \citenamefont {Peterson}, \citenamefont {Hu}, \citenamefont {Nowadnick},\
  and\ \citenamefont {Ding}}]{Hu2026TritonDFT}%
  \BibitemOpen
  \bibfield  {author} {\bibinfo {author} {\bibfnamefont {Z.}~\bibnamefont
  {Hu}}, \bibinfo {author} {\bibfnamefont {K.}~\bibnamefont {Talit}}, \bibinfo
  {author} {\bibfnamefont {Z.}~\bibnamefont {Wang}}, \bibinfo {author}
  {\bibfnamefont {H.}~\bibnamefont {Ahmad}}, \bibinfo {author} {\bibfnamefont
  {Y.}~\bibnamefont {Lin}}, \bibinfo {author} {\bibfnamefont {P.}~\bibnamefont
  {Kaur}}, \bibinfo {author} {\bibfnamefont {C.}~\bibnamefont {Lane}}, \bibinfo
  {author} {\bibfnamefont {E.~A.}\ \bibnamefont {Peterson}}, \bibinfo {author}
  {\bibfnamefont {Z.}~\bibnamefont {Hu}}, \bibinfo {author} {\bibfnamefont
  {E.~A.}\ \bibnamefont {Nowadnick}},\ and\ \bibinfo {author} {\bibfnamefont
  {Y.}~\bibnamefont {Ding}},\ }\href
  {https://doi.org/10.48550/arXiv.2603.03372} {\bibinfo {title} {{TritonDFT}:
  Automating {DFT} with a multi-agent framework}} (\bibinfo {year} {2026}),\
  \Eprint {https://arxiv.org/abs/2603.03372} {arXiv:2603.03372
  [cond-mat.mtrl-sci]} \BibitemShut {NoStop}%
\bibitem [{\citenamefont {Soleymanibrojeni}\ \emph {et~al.}(2026)\citenamefont
  {Soleymanibrojeni}, \citenamefont {Aydin}, \citenamefont {Guedes-Sobrinho},
  \citenamefont {Dias}, \citenamefont {Piotrowski}, \citenamefont {Wenzel},\
  and\ \citenamefont {R{\^e}go}}]{Soleymanibrojeni2026GENIUS}%
  \BibitemOpen
  \bibfield  {author} {\bibinfo {author} {\bibfnamefont {M.}~\bibnamefont
  {Soleymanibrojeni}}, \bibinfo {author} {\bibfnamefont {R.}~\bibnamefont
  {Aydin}}, \bibinfo {author} {\bibfnamefont {D.}~\bibnamefont
  {Guedes-Sobrinho}}, \bibinfo {author} {\bibfnamefont {A.~C.}\ \bibnamefont
  {Dias}}, \bibinfo {author} {\bibfnamefont {M.~J.}\ \bibnamefont
  {Piotrowski}}, \bibinfo {author} {\bibfnamefont {W.}~\bibnamefont {Wenzel}},\
  and\ \bibinfo {author} {\bibfnamefont {C.~R.~C.}\ \bibnamefont {R{\^e}go}},\
  }\bibfield  {title} {\bibinfo {title} {{GENIUS}: An agentic {AI} framework
  for autonomous design and execution of simulation protocols},\ }\href
  {https://doi.org/10.1038/s43246-026-01167-0} {\bibfield  {journal} {\bibinfo
  {journal} {Commun. Mater.}\ }\textbf {\bibinfo {volume} {7}},\ \bibinfo
  {pages} {115} (\bibinfo {year} {2026})}\BibitemShut {NoStop}%
\bibitem [{\citenamefont {Michaud-Rioux}\ \emph {et~al.}(2016)\citenamefont
  {Michaud-Rioux}, \citenamefont {Zhang},\ and\ \citenamefont
  {Guo}}]{MichaudRioux2016}%
  \BibitemOpen
  \bibfield  {author} {\bibinfo {author} {\bibfnamefont {V.}~\bibnamefont
  {Michaud-Rioux}}, \bibinfo {author} {\bibfnamefont {L.}~\bibnamefont
  {Zhang}},\ and\ \bibinfo {author} {\bibfnamefont {H.}~\bibnamefont {Guo}},\
  }\bibfield  {title} {\bibinfo {title} {{RESCU}: A real space electronic
  structure method},\ }\href {https://doi.org/10.1016/j.jcp.2015.12.014}
  {\bibfield  {journal} {\bibinfo  {journal} {J. Comput. Phys.}\ }\textbf
  {\bibinfo {volume} {307}},\ \bibinfo {pages} {593} (\bibinfo {year}
  {2016})}\BibitemShut {NoStop}%
\bibitem [{\citenamefont {Kresse}\ and\ \citenamefont
  {Furthm\"uller}(1996)}]{Kresse1996}%
  \BibitemOpen
  \bibfield  {author} {\bibinfo {author} {\bibfnamefont {G.}~\bibnamefont
  {Kresse}}\ and\ \bibinfo {author} {\bibfnamefont {J.}~\bibnamefont
  {Furthm\"uller}},\ }\bibfield  {title} {\bibinfo {title} {Efficient iterative
  schemes for {\it ab initio} total-energy calculations using a plane-wave
  basis set},\ }\href {https://doi.org/10.1103/PhysRevB.54.11169} {\bibfield
  {journal} {\bibinfo  {journal} {Phys. Rev. B}\ }\textbf {\bibinfo {volume}
  {54}},\ \bibinfo {pages} {11169} (\bibinfo {year} {1996})}\BibitemShut
  {NoStop}%
\bibitem [{\citenamefont {Jain}\ \emph {et~al.}(2013)\citenamefont {Jain},
  \citenamefont {Ong}, \citenamefont {Hautier}, \citenamefont {Chen},
  \citenamefont {Richards}, \citenamefont {Dacek}, \citenamefont {Cholia},
  \citenamefont {Gunter}, \citenamefont {Skinner}, \citenamefont {Ceder},\ and\
  \citenamefont {Persson}}]{Jain2013MaterialsProject}%
  \BibitemOpen
  \bibfield  {author} {\bibinfo {author} {\bibfnamefont {A.}~\bibnamefont
  {Jain}}, \bibinfo {author} {\bibfnamefont {S.~P.}\ \bibnamefont {Ong}},
  \bibinfo {author} {\bibfnamefont {G.}~\bibnamefont {Hautier}}, \bibinfo
  {author} {\bibfnamefont {W.}~\bibnamefont {Chen}}, \bibinfo {author}
  {\bibfnamefont {W.~D.}\ \bibnamefont {Richards}}, \bibinfo {author}
  {\bibfnamefont {S.}~\bibnamefont {Dacek}}, \bibinfo {author} {\bibfnamefont
  {S.}~\bibnamefont {Cholia}}, \bibinfo {author} {\bibfnamefont
  {D.}~\bibnamefont {Gunter}}, \bibinfo {author} {\bibfnamefont
  {D.}~\bibnamefont {Skinner}}, \bibinfo {author} {\bibfnamefont
  {G.}~\bibnamefont {Ceder}},\ and\ \bibinfo {author} {\bibfnamefont {K.~A.}\
  \bibnamefont {Persson}},\ }\bibfield  {title} {\bibinfo {title} {Commentary:
  The materials project: A materials genome approach to accelerating materials
  innovation},\ }\href {https://doi.org/10.1063/1.4812323} {\bibfield
  {journal} {\bibinfo  {journal} {APL Mater.}\ }\textbf {\bibinfo {volume}
  {1}},\ \bibinfo {pages} {011002} (\bibinfo {year} {2013})}\BibitemShut
  {NoStop}%
\bibitem [{\citenamefont {Li}\ and\ \citenamefont {Li}(2017)}]{Li2017}%
  \BibitemOpen
  \bibfield  {author} {\bibinfo {author} {\bibfnamefont {R.}~\bibnamefont
  {Li}}\ and\ \bibinfo {author} {\bibfnamefont {C.}~\bibnamefont {Li}},\
  }\bibfield  {title} {\bibinfo {title} {Photocatalytic water splitting on
  semiconductor-based photocatalysts},\ }in\ \href
  {https://doi.org/10.1016/bs.acat.2017.09.001} {\emph {\bibinfo {booktitle}
  {Advances in Catalysis}}},\ Vol.~\bibinfo {volume} {60}\ (\bibinfo
  {publisher} {Elsevier},\ \bibinfo {year} {2017})\ pp.\ \bibinfo {pages}
  {1--57}\BibitemShut {NoStop}%
\bibitem [{\citenamefont {Hisatomi}\ and\ \citenamefont
  {Domen}(2019)}]{Hisatomi2019}%
  \BibitemOpen
  \bibfield  {author} {\bibinfo {author} {\bibfnamefont {T.}~\bibnamefont
  {Hisatomi}}\ and\ \bibinfo {author} {\bibfnamefont {K.}~\bibnamefont
  {Domen}},\ }\bibfield  {title} {\bibinfo {title} {Reaction systems for solar
  hydrogen production via water splitting with particulate semiconductor
  photocatalysts},\ }\href {https://doi.org/10.1038/s41929-019-0242-6}
  {\bibfield  {journal} {\bibinfo  {journal} {Nat. Catal.}\ }\textbf {\bibinfo
  {volume} {2}},\ \bibinfo {pages} {387} (\bibinfo {year} {2019})}\BibitemShut
  {NoStop}%
\bibitem [{\citenamefont {Vurgaftman}\ \emph {et~al.}(2001)\citenamefont
  {Vurgaftman}, \citenamefont {Meyer},\ and\ \citenamefont
  {Ram-Mohan}}]{Vurgaftman2001}%
  \BibitemOpen
  \bibfield  {author} {\bibinfo {author} {\bibfnamefont {I.}~\bibnamefont
  {Vurgaftman}}, \bibinfo {author} {\bibfnamefont {J.~R.}\ \bibnamefont
  {Meyer}},\ and\ \bibinfo {author} {\bibfnamefont {L.~R.}\ \bibnamefont
  {Ram-Mohan}},\ }\bibfield  {title} {\bibinfo {title} {Band parameters for
  {III}--{V} compound semiconductors and their alloys},\ }\href
  {https://doi.org/10.1063/1.1368156} {\bibfield  {journal} {\bibinfo
  {journal} {J. Appl. Phys.}\ }\textbf {\bibinfo {volume} {89}},\ \bibinfo
  {pages} {5815} (\bibinfo {year} {2001})}\BibitemShut {NoStop}%
\bibitem [{\citenamefont {Wang}\ \emph {et~al.}(2019)\citenamefont {Wang},
  \citenamefont {Wu}, \citenamefont {Sun},\ and\ \citenamefont
  {Mi}}]{Wang2019}%
  \BibitemOpen
  \bibfield  {author} {\bibinfo {author} {\bibfnamefont {Y.}~\bibnamefont
  {Wang}}, \bibinfo {author} {\bibfnamefont {Y.}~\bibnamefont {Wu}}, \bibinfo
  {author} {\bibfnamefont {K.}~\bibnamefont {Sun}},\ and\ \bibinfo {author}
  {\bibfnamefont {Z.}~\bibnamefont {Mi}},\ }\bibfield  {title} {\bibinfo
  {title} {A quadruple-band metal-nitride nanowire artificial photosynthesis
  system for high efficiency photocatalytic overall solar water splitting},\
  }\href {https://doi.org/10.1039/C9MH00257J} {\bibfield  {journal} {\bibinfo
  {journal} {Mater. Horiz.}\ }\textbf {\bibinfo {volume} {6}},\ \bibinfo
  {pages} {1454} (\bibinfo {year} {2019})}\BibitemShut {NoStop}%
\bibitem [{\citenamefont {Zhou}\ \emph {et~al.}(2023)\citenamefont {Zhou},
  \citenamefont {Navid}, \citenamefont {Ma}, \citenamefont {Xiao},
  \citenamefont {Wang}, \citenamefont {Ye}, \citenamefont {Zhou}, \citenamefont
  {Sun},\ and\ \citenamefont {Mi}}]{Zhou2023}%
  \BibitemOpen
  \bibfield  {author} {\bibinfo {author} {\bibfnamefont {P.}~\bibnamefont
  {Zhou}}, \bibinfo {author} {\bibfnamefont {I.~A.}\ \bibnamefont {Navid}},
  \bibinfo {author} {\bibfnamefont {Y.}~\bibnamefont {Ma}}, \bibinfo {author}
  {\bibfnamefont {Y.}~\bibnamefont {Xiao}}, \bibinfo {author} {\bibfnamefont
  {P.}~\bibnamefont {Wang}}, \bibinfo {author} {\bibfnamefont {Z.}~\bibnamefont
  {Ye}}, \bibinfo {author} {\bibfnamefont {B.}~\bibnamefont {Zhou}}, \bibinfo
  {author} {\bibfnamefont {K.}~\bibnamefont {Sun}},\ and\ \bibinfo {author}
  {\bibfnamefont {Z.}~\bibnamefont {Mi}},\ }\bibfield  {title} {\bibinfo
  {title} {Solar-to-hydrogen efficiency of more than 9\% in photocatalytic
  water splitting},\ }\href {https://doi.org/10.1038/s41586-022-05399-1}
  {\bibfield  {journal} {\bibinfo  {journal} {Nature}\ }\textbf {\bibinfo
  {volume} {613}},\ \bibinfo {pages} {66} (\bibinfo {year} {2023})}\BibitemShut
  {NoStop}%
\bibitem [{\citenamefont {Kamat}(2012)}]{Kamat2012}%
  \BibitemOpen
  \bibfield  {author} {\bibinfo {author} {\bibfnamefont {P.~V.}\ \bibnamefont
  {Kamat}},\ }\bibfield  {title} {\bibinfo {title} {Manipulation of charge
  transfer across semiconductor interface. {A} criterion that cannot be ignored
  in photocatalyst design},\ }\href {https://doi.org/10.1021/jz201629p}
  {\bibfield  {journal} {\bibinfo  {journal} {J. Phys. Chem. Lett.}\ }\textbf
  {\bibinfo {volume} {3}},\ \bibinfo {pages} {663} (\bibinfo {year}
  {2012})}\BibitemShut {NoStop}%
\bibitem [{\citenamefont {Tersoff}(1984)}]{Tersoff1984}%
  \BibitemOpen
  \bibfield  {author} {\bibinfo {author} {\bibfnamefont {J.}~\bibnamefont
  {Tersoff}},\ }\bibfield  {title} {\bibinfo {title} {Schottky barrier heights
  and the continuum of gap states},\ }\href
  {https://doi.org/10.1103/PhysRevLett.52.465} {\bibfield  {journal} {\bibinfo
  {journal} {Phys. Rev. Lett.}\ }\textbf {\bibinfo {volume} {52}},\ \bibinfo
  {pages} {465} (\bibinfo {year} {1984})}\BibitemShut {NoStop}%
\bibitem [{\citenamefont {Liu}\ \emph {et~al.}(2024)\citenamefont {Liu},
  \citenamefont {Wang},\ and\ \citenamefont {Pinna}}]{Liu2024}%
  \BibitemOpen
  \bibfield  {author} {\bibinfo {author} {\bibfnamefont {Y.}~\bibnamefont
  {Liu}}, \bibinfo {author} {\bibfnamefont {Y.}~\bibnamefont {Wang}},\ and\
  \bibinfo {author} {\bibfnamefont {N.}~\bibnamefont {Pinna}},\ }\bibfield
  {title} {\bibinfo {title} {Atomically precise metal nanoclusters for
  photocatalytic water splitting},\ }\href
  {https://doi.org/10.1021/acsmaterialslett.4c00622} {\bibfield  {journal}
  {\bibinfo  {journal} {ACS Mater. Lett.}\ }\textbf {\bibinfo {volume} {6}},\
  \bibinfo {pages} {2995} (\bibinfo {year} {2024})}\BibitemShut {NoStop}%
\bibitem [{\citenamefont {Yuan}\ \emph {et~al.}(2026)\citenamefont {Yuan},
  \citenamefont {Andersson}, \citenamefont {Metha}, \citenamefont {Mi},\ and\
  \citenamefont {Guo}}]{Yuan2026}%
  \BibitemOpen
  \bibfield  {author} {\bibinfo {author} {\bibfnamefont {S.}~\bibnamefont
  {Yuan}}, \bibinfo {author} {\bibfnamefont {G.~G.}\ \bibnamefont {Andersson}},
  \bibinfo {author} {\bibfnamefont {G.~F.}\ \bibnamefont {Metha}}, \bibinfo
  {author} {\bibfnamefont {Z.}~\bibnamefont {Mi}},\ and\ \bibinfo {author}
  {\bibfnamefont {H.}~\bibnamefont {Guo}},\ }\bibfield  {title} {\bibinfo
  {title} {Design principles for {III}-nitride-nanocluster photocatalysts from
  region-resolved electronic structure},\ }\href
  {https://arxiv.org/abs/2602.08113} {\bibfield  {journal} {\bibinfo  {journal}
  {arXiv}\ } (\bibinfo {year} {2026})},\ \Eprint
  {https://arxiv.org/abs/2602.08113} {arXiv:2602.08113 [cond-mat.mtrl-sci]}
  \BibitemShut {NoStop}%
\bibitem [{\citenamefont {Zunger}\ \emph {et~al.}(1990)\citenamefont {Zunger},
  \citenamefont {Wei}, \citenamefont {Ferreira},\ and\ \citenamefont
  {Bernard}}]{Zunger1990}%
  \BibitemOpen
  \bibfield  {author} {\bibinfo {author} {\bibfnamefont {A.}~\bibnamefont
  {Zunger}}, \bibinfo {author} {\bibfnamefont {S.-H.}\ \bibnamefont {Wei}},
  \bibinfo {author} {\bibfnamefont {L.~G.}\ \bibnamefont {Ferreira}},\ and\
  \bibinfo {author} {\bibfnamefont {J.~E.}\ \bibnamefont {Bernard}},\
  }\bibfield  {title} {\bibinfo {title} {Special quasirandom structures},\
  }\href {https://doi.org/10.1103/PhysRevLett.65.353} {\bibfield  {journal}
  {\bibinfo  {journal} {Phys. Rev. Lett.}\ }\textbf {\bibinfo {volume} {65}},\
  \bibinfo {pages} {353} (\bibinfo {year} {1990})}\BibitemShut {NoStop}%
\bibitem [{\citenamefont {Batatia}\ \emph {et~al.}(2025)\citenamefont
  {Batatia}, \citenamefont {Benner}, \citenamefont {Chiang}, \citenamefont
  {Elena}, \citenamefont {Kov{\'a}cs}, \citenamefont {Riebesell} \emph
  {et~al.}}]{Batatia2025}%
  \BibitemOpen
  \bibfield  {author} {\bibinfo {author} {\bibfnamefont {I.}~\bibnamefont
  {Batatia}}, \bibinfo {author} {\bibfnamefont {P.}~\bibnamefont {Benner}},
  \bibinfo {author} {\bibfnamefont {Y.}~\bibnamefont {Chiang}}, \bibinfo
  {author} {\bibfnamefont {A.~M.}\ \bibnamefont {Elena}}, \bibinfo {author}
  {\bibfnamefont {D.~P.}\ \bibnamefont {Kov{\'a}cs}}, \bibinfo {author}
  {\bibfnamefont {J.}~\bibnamefont {Riebesell}}, \emph {et~al.},\ }\bibfield
  {title} {\bibinfo {title} {A foundation model for atomistic materials
  chemistry},\ }\href {https://doi.org/10.1063/5.0297006} {\bibfield  {journal}
  {\bibinfo  {journal} {J. Chem. Phys.}\ }\textbf {\bibinfo {volume} {163}},\
  \bibinfo {pages} {184110} (\bibinfo {year} {2025})}\BibitemShut {NoStop}%
\bibitem [{\citenamefont {Perdew}\ \emph {et~al.}(1996)\citenamefont {Perdew},
  \citenamefont {Burke},\ and\ \citenamefont {Ernzerhof}}]{Perdew1996}%
  \BibitemOpen
  \bibfield  {author} {\bibinfo {author} {\bibfnamefont {J.~P.}\ \bibnamefont
  {Perdew}}, \bibinfo {author} {\bibfnamefont {K.}~\bibnamefont {Burke}},\ and\
  \bibinfo {author} {\bibfnamefont {M.}~\bibnamefont {Ernzerhof}},\ }\bibfield
  {title} {\bibinfo {title} {Generalized gradient approximation made simple},\
  }\href {https://doi.org/10.1103/PhysRevLett.77.3865} {\bibfield  {journal}
  {\bibinfo  {journal} {Phys. Rev. Lett.}\ }\textbf {\bibinfo {volume} {77}},\
  \bibinfo {pages} {3865} (\bibinfo {year} {1996})}\BibitemShut {NoStop}%
\bibitem [{\citenamefont {Vegh}\ \emph {et~al.}(2026)\citenamefont {Vegh},
  \citenamefont {Zhu}, \citenamefont {Prentki}, \citenamefont {Beaudoin},
  \citenamefont {Yuan},\ and\ \citenamefont {Guo}}]{Vegh2026LatticeMindData}%
  \BibitemOpen
  \bibfield  {author} {\bibinfo {author} {\bibfnamefont {N.~M.}\ \bibnamefont
  {Vegh}}, \bibinfo {author} {\bibfnamefont {Y.}~\bibnamefont {Zhu}}, \bibinfo
  {author} {\bibfnamefont {R.~J.}\ \bibnamefont {Prentki}}, \bibinfo {author}
  {\bibfnamefont {F.}~\bibnamefont {Beaudoin}}, \bibinfo {author}
  {\bibfnamefont {S.}~\bibnamefont {Yuan}},\ and\ \bibinfo {author}
  {\bibfnamefont {H.}~\bibnamefont {Guo}},\ }\href
  {https://github.com/nathaniel-vegh/latticemind-data-package} {\bibinfo
  {title} {{LatticeMind} benchmark data}},\ \bibinfo {howpublished} {GitHub
  repository} (\bibinfo {year} {2026}),\ \bibinfo {note} {accessed July 14,
  2026}\BibitemShut {NoStop}%
\end{thebibliography}%


\begin{thebibliography}{0}%
\makeatletter
\providecommand \@ifxundefined [1]{%
 \@ifx{#1\undefined}
}%
\providecommand \@ifnum [1]{%
 \ifnum #1\expandafter \@firstoftwo
 \else \expandafter \@secondoftwo
 \fi
}%
\providecommand \@ifx [1]{%
 \ifx #1\expandafter \@firstoftwo
 \else \expandafter \@secondoftwo
 \fi
}%
\providecommand \natexlab [1]{#1}%
\providecommand \enquote  [1]{``#1''}%
\providecommand \bibnamefont  [1]{#1}%
\providecommand \bibfnamefont [1]{#1}%
\providecommand \citenamefont [1]{#1}%
\providecommand \href@noop [0]{\@secondoftwo}%
\providecommand \href [0]{\begingroup \@sanitize@url \@href}%
\providecommand \@href[1]{\@@startlink{#1}\@@href}%
\providecommand \@@href[1]{\endgroup#1\@@endlink}%
\providecommand \@sanitize@url [0]{\catcode `\\12\catcode `\$12\catcode
  `\&12\catcode `\#12\catcode `\^12\catcode `\_12\catcode `\%12\relax}%
\providecommand \@@startlink[1]{}%
\providecommand \@@endlink[0]{}%
\providecommand \url  [0]{\begingroup\@sanitize@url \@url }%
\providecommand \@url [1]{\endgroup\@href {#1}{\urlprefix }}%
\providecommand \urlprefix  [0]{URL }%
\providecommand \Eprint [0]{\href }%
\providecommand \doibase [0]{https://doi.org/}%
\providecommand \selectlanguage [0]{\@gobble}%
\providecommand \bibinfo  [0]{\@secondoftwo}%
\providecommand \bibfield  [0]{\@secondoftwo}%
\providecommand \translation [1]{[#1]}%
\providecommand \BibitemOpen [0]{}%
\providecommand \bibitemStop [0]{}%
\providecommand \bibitemNoStop [0]{.\EOS\space}%
\providecommand \EOS [0]{\spacefactor3000\relax}%
\providecommand \BibitemShut  [1]{\csname bibitem#1\endcsname}%
\let\auto@bib@innerbib\@empty
\end{thebibliography}%

\end{document}


\title{Supplemental Material for LatticeMind: A Validation-Gated Agentic Framework for Density Functional Theory Workflows}

\maketitle

\setcounter{table}{0}
\setcounter{figure}{0}
\setcounter{section}{0}
\setcounter{equation}{0}
\renewcommand{\thetable}{S\arabic{table}}
\renewcommand{\thefigure}{S\arabic{figure}}
\renewcommand{\thesection}{S\arabic{section}}
\renewcommand{\theequation}{S\arabic{equation}}

\section{Completed benchmark scope}

The manuscript reports a completed 44-case Real-Space Electronic Structure CalcUlator (RESCU) benchmark scored from frozen case prompts, LatticeMind run directories, solver outputs, validation logs, scoring records, and objective reference files. Here, a case means a fixed natural-language request together with the files and checks needed to score it; a backend means the language model used behind the same LatticeMind workflow engine. Each case was run once with the same LatticeMind architecture using GPT-5.4-mini, Gemini 3 Flash Preview, or Claude Sonnet 4.6, yielding 132 scored LatticeMind runs.

The supplemental analysis also includes a matched architecture ablation, a component-removal study on a fixed 23-case subset with three repeats per variant. This test isolates the contribution of workflow contracts, documentation retrieval, deterministic validation, and repair while using the same external scoring policy.

\section{Systems and run provenance}

\begin{table*}[!htbp]
\caption{\label{tab:si_system_manifest}
Systems used in the completed benchmark. ``Project tools'' denotes the LatticeMind components that track project state, create files, validate inputs, route repairs, and package files for execution. Dates are run or scoring completion dates recorded in the source filenames and campaign metadata.}
\small
\begin{tabular}{p{0.24\textwidth}p{0.68\textwidth}}
\toprule
System & Model and shared configuration notes \\
\midrule
LatticeMind + GPT-5.4-mini & OpenAI \texttt{gpt-5.4-mini}; completed 2026-05-07; project tools, local RESCU documentation, deterministic validation, bounded repair, no web search; OpenAI reasoning effort ``quick'' \\
LatticeMind + Gemini 3 Flash Preview & Google \texttt{gemini-3-flash-preview}; completed 2026-05-08; same LatticeMind experimental setup; temperature 0.1 and topP 0.5 where accepted by the provider \\
LatticeMind + Claude Sonnet 4.6 & Anthropic \texttt{claude-sonnet-4-6}; completed 2026-05-08; same LatticeMind experimental setup; provider-default decoding settings \\
\bottomrule
\end{tabular}
\end{table*}

For the three LatticeMind benchmark systems, local documentation retrieval, workflow contracts, deterministic validation, bounded repair, and web-search exclusion were enabled under the same experimental protocol. The reported benchmark metrics are computed from the June 2026 uniform rescore, partitioned by system from the completed run directories and using the same scoring policy for all three backends. The rescore re-executed the scoring checks over the completed May 2026 run directories, checking every backend against one frozen reference set and tolerance with no per-model leniency; the prompts, frozen per-case scoring specifications, and completed run directories themselves were not modified. The ablation metrics are computed from the matched 552-run ablation records and the corresponding summary metrics.

\section{Benchmark case inventory}\label{si:cases}

\begin{table}[t]
\caption{\label{tab:si_family_manifest}
Family composition of the 44-case benchmark. SCF denotes self-consistent field, DOS denotes density of states, and PDOS denotes projected density of states.}
\begin{tabular}{lc}
\toprule
Task family & Cases \\
\midrule
Total-energy / SCF workflows & 19 \\
DOS / PDOS workflows & 12 \\
Band-structure workflows & 6 \\
Magnetic ground-state workflows & 2 \\
Phonon workflows & 2 \\
Potential workflow & 1 \\
Structure-relaxation workflow & 1 \\
Band-edge trend workflow & 1 \\
\bottomrule
\end{tabular}
\end{table}

Each case contains a natural-language prompt, allowed starting files, expected calculation steps, required files, and automatic checks. The checks include calculation completion, density reuse, workflow-contract satisfaction, selected structural/electronic scalar checks, and task-specific file checks. All 44 cases in the automatic scoring pass had frozen expected outputs and required no human expert adjudication. The prompts, case contracts, allowed starting files, and checks were frozen before the reported model runs; generated outputs were not used to alter case-specific criteria for those reported runs. Table~\ref{tab:si_case_inventory} lists every case by identifier and short description; the complete prompts and contracts are provided in the benchmark-data package.

\begin{table*}[!htbp]
\caption{\label{tab:si_case_inventory}
Per-case inventory of the 44-case RESCU benchmark. Each row corresponds to one frozen natural-language prompt with an associated workflow contract, allowed starting files, and automatic checks. Full prompts and contracts are provided in the benchmark-data package.}
\small
\setlength{\tabcolsep}{4pt}
\renewcommand{\arraystretch}{0.88}
\begin{tabular}{@{}p{0.31\textwidth}p{0.63\textwidth}@{}}
\toprule
Case identifier & Short description \\
\midrule
\path{al_dfpt_gamma_phonon} & Aluminum Gamma DFPT phonon smoke test \\
\path{al_kmesh_scan} & Three-point aluminum k-mesh scan \\
\path{aln_wurtzite_scf_band} & Wurtzite AlN SCF plus band structure \\
\path{alpha_al2o3_primitive_scf} & Primitive alpha-Al$_2$O$_3$ corundum SCF \\
\path{alpha_al2o3_scf_dos} & Alpha-Al$_2$O$_3$ (sapphire) SCF plus DOS \\
\path{bn_zincblende_scf_band} & Zincblende BN SCF plus band structure \\
\path{carbon_scf_plus_potential} & Carbon SCF plus potential \\
\path{catio3_perovskite_scf} & CaTiO$_3$ perovskite reference-cell SCF \\
\path{cdse_zincblende_scf_dos} & Zinc-blende CdSe SCF plus DOS \\
\path{diamond_band_compressed_cell} & Compressed-cell diamond SCF plus band structure \\
\path{diamond_eos_three_point} & Three-point diamond EOS study \\
\path{diamond_eos_three_point_compact} & Compact three-point diamond EOS study \\
\path{diamond_scf_band_fixed_cell} & Fixed-cell diamond SCF plus band structure \\
\path{diamond_scf_reference_cell} & Diamond reference-cell SCF \\
\path{diamond_two_strain_band_gap} & Two-strain diamond SCF plus band comparison \\
\path{fe_spin_dos} & Spin-polarized iron SCF plus DOS (open-ended prompt; magnetic setup and file layout left to the agent) \\
\path{fe_spin_scf_dos_chain} & Spin-polarized iron SCF plus DOS (fully specified chain with explicit Fe moments and exact input paths) \\
\path{gaas_scf_band_structure} & Zinc-blende GaAs SCF plus band structure \\
\path{ge_scf_primitive_cell} & Primitive-cell SCF for bulk germanium \\
\path{graphene_band_dos_chain} & Graphene SCF plus band structure and DOS \\
\path{graphene_kmesh_scan} & Deterministic three-point graphene k-mesh scan \\
\path{graphene_pdos_pi_window} & Graphene SCF plus PDOS near the Fermi level \\
\path{graphene_scf_vacuum_cell} & Graphene SCF in a small vacuum cell \\
\path{hfo2_monoclinic_scf} & Monoclinic HfO$_2$ high-k oxide SCF \\
\path{inse_layered_scf} & Layered InSe reference-cell SCF \\
\path{licl_rocksalt_scf} & LiCl rocksalt reference-cell SCF \\
\path{mgb2_scf_dos} & MgB$_2$ SCF plus DOS \\
\path{mgo_eos_three_point} & Three-point MgO EOS study \\
\path{nacl_rocksalt_scf} & NaCl rocksalt reference-cell SCF \\
\path{nb_bcc_scf_dos} & BCC niobium SCF plus DOS \\
\path{nbse2_layered_scf_dos} & Layered NbSe$_2$ SCF plus DOS \\
\path{nio_magnetic_oxide_scf} & NiO magnetic-oxide proxy SCF \\
\path{o2_noncollinear_spin_smoke} & O$_2$ noncollinear spin smoke SCF \\
\path{pbse_rocksalt_scf_dos} & Rocksalt PbSe SCF plus DOS \\
\path{pbtio3_tetragonal_scf} & Tetragonal PbTiO$_3$ ferroelectric SCF \\
\path{si_finite_displacement_phonon} & Silicon finite-displacement phonon dispersion \\
\path{si_kmesh_scan} & Small k-mesh convergence scan for silicon \\
\path{si_lcao_scf_primitive} & Silicon primitive-cell LCAO SCF \\
\path{si_scf_dos_fixed_cell} & Fixed-cell silicon SCF plus DOS \\
\path{sic_zincblende_scf_band} & Zincblende 3C-SiC SCF plus band structure \\
\path{sige_ordered8_relax} & Ordered 8-atom Si$_{0.75}$Ge$_{0.25}$ relaxation \\
\path{sige_ordered8_scf_dos} & Ordered Si$_{0.75}$Ge$_{0.25}$ SCF plus DOS \\
\path{sio2_alpha_quartz_scf} & Alpha-quartz SiO$_2$ reference-cell SCF \\
\path{ta_bcc_reference_cell} & BCC tantalum reference-cell SCF \\
\bottomrule
\end{tabular}
\end{table*}

\section{Run procedure and scoring}\label{si:scoring}

Each main benchmark run used the same procedure:

\begin{enumerate}
\item Start from a clean benchmark project directory.
\item Submit the frozen case prompt to LatticeMind with the selected model backend.
\item Allow LatticeMind to search the local RESCU documentation corpus, plan the workflow, choose calculation settings, build inputs, validate, repair if needed, and stop when the files needed for scoring are available or when validation prevents further execution.
\item Package generated inputs, solver outputs, validation records, run logs, and scoring files.
\item Score the completed run directories with the frozen deterministic scoring script and existing RESCU outputs.
\end{enumerate}

For each run, the scorer writes a deterministic JavaScript Object Notation (JSON) file containing generated-file counts, file-completion markers, predefined-check outcomes, and status flags from the run log. The scoring script is outside the agent generation loop and does not call a language model. Execution approval was implemented as a fixed policy: allowed low-cost RESCU and post-processing steps were automatically approved, disallowed actions were blocked, and no case-specific manual file edits were made before scoring.

For RESCU, \texttt{rho.in} is the input reference used to reuse a previously converged electron density in a downstream calculation.

\begin{table*}[!htbp]
\caption{\label{tab:si_contract_example}
Example workflow-contract fields for a representative SCF-to-DOS benchmark case.}
\small
\begin{tabular}{p{0.24\textwidth}p{0.64\textwidth}}
\toprule
Contract field & Example value \\
\midrule
Task family & SCF $\rightarrow$ DOS / PDOS \\
Required upstream file & Converged SCF save directory containing the density reused by the downstream calculation \\
Required generated input & DOS input file with a \texttt{rho.in} dependency and any requested projection block \\
Allowed execution action & Run RESCU DOS calculation only after the upstream density path exists \\
Validator check & Confirm that \texttt{rho.in} points to an existing SCF output and that unsupported or ambiguous keywords are absent \\
Reference check & Confirm that the generated files include the requested DOS/PDOS output and satisfy the case-specific dependency and observable checks \\
\bottomrule
\end{tabular}
\end{table*}

\section{Workflow-level reasoning evidence}

The benchmark evaluates whether an agent-generated project preserves the workflow dependencies and solver conventions required for a valid first-principles calculation. Table~\ref{tab:si_reasoning_evidence} summarizes the reasoning classes that appear repeatedly in the benchmark contracts, validator findings, repair logs, and setup-reconstruction checks. PLDOS denotes projected local density of states, and DFPT denotes density-functional perturbation theory. Supporting scoring and ablation records are included in the benchmark-data package.

\begin{table*}[!htbp]
\caption{\label{tab:si_reasoning_evidence}
Workflow-level reasoning evidence used to interpret LatticeMind's successful runs.}
\small
\begin{tabular}{p{0.20\textwidth}p{0.70\textwidth}}
\toprule
Evidence class & What is checked and why it matters \\
\midrule
Dependency reasoning & Continuation tasks must reuse the intended SCF density for DOS, band, PDOS, PLDOS, potential, or DFPT stages. LatticeMind checks the required dependency edges, active input files, and \texttt{rho.in} handoffs so that follow-up calculations are not run on a missing or unrelated density. \\
Scan reasoning & Parameter, structure, strain, field, and functional scans must create distinct calculation instances and non-overwriting output locations. The workflow contract checks scan variants, unique save paths, and graph completeness to avoid collapsed scans or overwritten results. \\
Solver-convention reasoning & Generated RESCU inputs must use supported fields, legal aliases, valid value types, and calculation-family-specific settings. Deterministic field and value validators catch inputs that look plausible but would be invalid for the solver. \\
Projection reasoning & PDOS and PLDOS requests must map projected quantities to explicit species, orbital channels, atom lists, layers, or surface regions. Projection labels, generated helper scripts, and generated-file checks make the projected quantities traceable and comparable. \\
Repair reasoning & Validator feedback should regenerate the responsible file or input block before execution proceeds. The limited repair loop prevents execution after a known contract or input-rule failure. \\
\bottomrule
\end{tabular}
\end{table*}

\begin{table*}[!htbp]
\caption{\label{tab:si_metric_defs}
Operational definitions for the benchmark metrics used in the manuscript and this Supplemental Material.}
\small
\begin{tabular}{p{0.20\textwidth}p{0.43\textwidth}p{0.27\textwidth}}
\toprule
Metric & Definition & Typical failure example \\
\midrule
Output produced & Completed benchmark folder contains RESCU or post-processing output from an approved execution step & Inputs exist, but no solver or post-processing output marker exists \\
Workflow complete & Run satisfies the case workflow contract and required file checks & Missing downstream density reuse, projection block, band path, expected output file, or requested follow-up stage \\
Passed checks & Run passes the predefined automatic checks for the case & Executed or generated file has an incorrect scalar, dependency, structure, or task-specific observable \\
Passed before repair & Run reaches execution readiness without an additional repair loop & Not applicable; readiness category for successful runs \\
Passed after repair & Run reaches execution readiness after at least one repair loop & Initial validation fails but bounded repair produces a valid workflow \\
\bottomrule
\end{tabular}
\end{table*}

\section{Aggregate benchmark results}\label{si:aggregate}

Table~\ref{tab:si_benchmark_aggregate} reports aggregate automatic scoring metrics for each backend. All three backends passed the predefined checks for all 44 benchmark cases. Because the family-resolved pass counts are therefore all complete, the family breakdown is reported only as the benchmark case inventory in Table~\ref{tab:si_family_manifest}; per-case records are provided in the benchmark-data package.

The 95\% intervals are Wilson binomial confidence intervals for the pass probability, computed from the listed pass/fail counts; for 44/44 passes, this gives 92.0--100\%.

\begin{table*}[!htbp]
\caption{\label{tab:si_benchmark_aggregate}
Aggregate automatic scoring results for the completed 44-case benchmark. The 95\% interval is the Wilson confidence interval for the pass probability inferred from 44 pass/fail cases.}
\begin{tabular}{lccccc}
\toprule
System & Output produced & Workflow complete & Passed checks & Check-pass 95\% interval & Mean score \\
\midrule
GPT-5.4-mini & 44/44 (100.0\%) & 44/44 (100.0\%) & 44/44 (100.0\%) & 92.0--100\% & 1.000 \\
Gemini 3 Flash Preview & 44/44 (100.0\%) & 44/44 (100.0\%) & 44/44 (100.0\%) & 92.0--100\% & 1.000 \\
Claude Sonnet 4.6 & 44/44 (100.0\%) & 44/44 (100.0\%) & 44/44 (100.0\%) & 92.0--100\% & 1.000 \\
\bottomrule
\end{tabular}
\end{table*}

Across the 44-case benchmark, mean agent token usage per run was 59k for GPT-5.4-mini, 65k for Claude Sonnet 4.6, and 155k for Gemini 3 Flash Preview (Table~\ref{tab:si_token_cost}); medians are lower, reflecting a right-skewed distribution from a few multi-instance scan and repaired cases. The two lowest-cost backends reach the same complete reference-pass rate, consistent with the main-text finding that reliability here is set by the workflow architecture rather than the per-run token budget.

\begin{table}[t]
\caption{\label{tab:si_token_cost}
Agent token usage per run over the 44-case benchmark. Because every case passed, per-run and per-successful-run costs coincide.}
\small
\begin{tabular}{lcc}
\toprule
Backend & Mean & Median \\
\midrule
GPT-5.4-mini & 59k & 51k \\
Claude Sonnet 4.6 & 65k & 55k \\
Gemini 3 Flash Preview & 155k & 124k \\
\bottomrule
\end{tabular}
\end{table}

\section{Recovery summary}

Under the same June 2026 automatic scorer, all 44 cases passed the reference checks on all three backends. The remaining variation is whether the validation-gated repair loop was invoked before a case passed, as shown in main-text Fig.~5(a,c). Across the 132 runs, 119 (90\%) passed on the first validated plan and 13 (10\%) were recovered after a check caught a problem; no run required more than one repair cycle.

GPT-5.4-mini recovered three cases by one repair cycle and passed the remaining 41 on the first validated plan. Gemini 3 Flash Preview and Claude Sonnet 4.6 each recovered five cases by one repair cycle and passed the remaining 39 on the first plan. The repaired cases were concentrated in magnetic, ordered-alloy, $\alpha$-quartz, and finite-displacement-phonon workflows; per-case identities are included in the benchmark-data package.

\section{Matched architecture ablation}\label{si:ablation}

The ablation study used a fixed 23-case subset (52\% of the 44-case benchmark) spanning simple SCF calculations, oxides and layered materials, SCF-to-DOS and SCF-to-band density handoffs, $k$-mesh and equation-of-state scans, alloy relaxation, magnetism, and phonon workflows, with three repeats per case. The subset was chosen to span the benchmark's task families while remaining weighted toward the harder multistage workflows, so it is deliberately more demanding than the full benchmark; consequently the recovery contribution measured here is larger than on the full 44-case set. A density handoff means that a downstream calculation reuses a previously converged electron density rather than starting from an unrelated file.

The study compares eight architecture variants: the full LatticeMind agent; four single-component ablations (no documentation, no workflow contract, no agent validator, and no repair loop); a reactive trial-and-repair baseline that repairs only after failed execution; and two direct-LLM baselines with and without documentation. The full agent used local documentation, a workflow contract, an agent-side validator, and at most three validation-triggered repair attempts before execution. Every variant was run on all 23 cases with three repeats under the same model, external scoring script, and predefined checks.

\begin{figure*}[!htbp]
\centering
\includegraphics[width=\FigSIAblationCaseMatrixWidth]{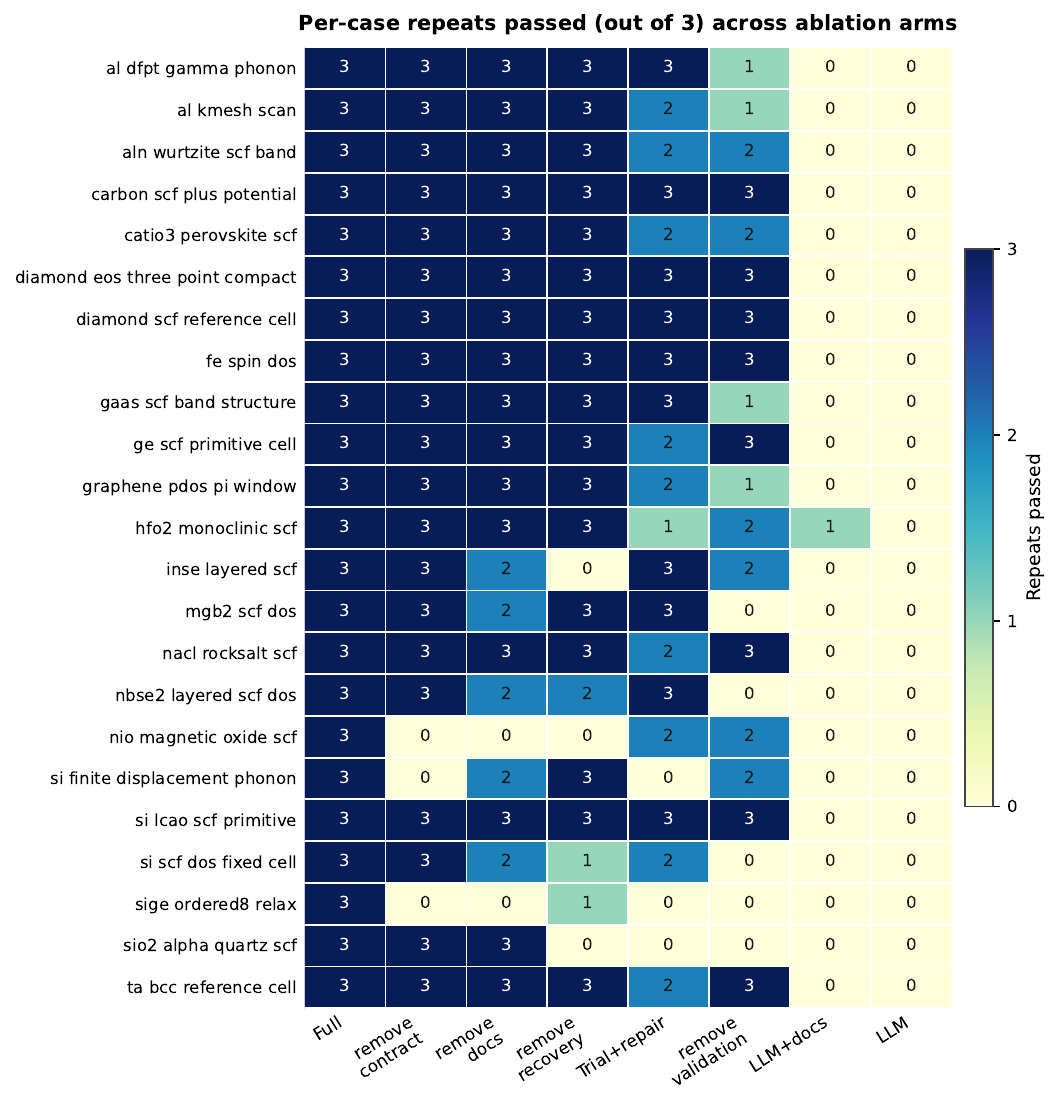}
\caption{\label{fig:si_ablation_case_matrix}
Per-case outcome matrix for the matched ablation. Each cell gives the number of repeats (out of three) that passed the reference checks, for each case (rows) and architecture variant (columns). Darker blue-green cells mark more repeats passing; pale cells mark zero passing repeats.}
\end{figure*}

For each ablation arm, the primary outcome is whether the generated project produced the expected files and matched all predefined numerical and file-based checks. Over 69 runs per arm, the reference-check pass counts were: full LatticeMind, 69/69; no workflow contract, 60/69; no documentation, 58/69; no repair loop, 55/69; trial-and-repair, 49/69; no agent validator, 40/69; direct LLM with documentation, 1/69; and direct LLM without documentation, 0/69. Operational completion followed the same ordering: full LatticeMind, 100.0\%; no workflow contract, 91.3\%; no documentation, 88.4\%; no repair loop, 84.1\%; trial-and-repair, 76.8\%; no agent validator, 63.8\%; direct LLM with documentation, 1.4\%; and direct LLM without documentation, 0.0\%. The full agent used 49k tokens per passing run, whereas the trial-and-repair baseline used 102k and the direct-LLM-with-documentation baseline used 574k because only one run passed.

In this controlled comparison, removing documentation retrieval and removing the workflow contract produced comparable leave-one-out losses of 15.9 and 13.0 percentage points, respectively, whereas removing the agent-side validator produced the largest loss of 42.0 percentage points. The cumulative build-up separately shows that adding the workflow contract produced the largest incremental gain.

Per-case ablation summaries are included in the benchmark-data package.

\section{Solver-agnostic VASP generalization}\label{si:vasp}

To test whether the LatticeMind architecture is specific to RESCU, we applied it to the Vienna Ab initio Simulation Package (VASP), changing only the code-specific configuration: allowed input fields and validation rules, documentation source, and helper scripts. The 19-case benchmark subset spans single-point SCF calculations, multi-instance $k$-mesh and equation-of-state scans, and (the majority of the set) SCF-to-DOS and SCF-to-band chains and a spin-polarized SCF-to-DOS chain across metals (Nb, MgB$_2$), a magnetic solid (Fe), wide-gap insulators ($\alpha$-Al$_2$O$_3$, cubic BN, diamond), and wurtzite semiconductors (ZnO, GaN). Using the same GPT-5.4-mini backend, LatticeMind passed the reference checks in all 19 cases (Table~\ref{tab:si_vasp_cases}). For every multistage case we further confirmed that the converged lattice constant and band gap fall within reviewed Materials Project and PBE reference windows (metals gapless, and PBE gaps of 4.5, 2.0, and 0.8\,eV for cubic BN, GaN, and ZnO), so the chains are physically correct, not merely runnable.

\begin{table}[!htbp]
\caption{\label{tab:si_vasp_cases}
Per-case inventory for the 19-case VASP transfer benchmark. Each row is one frozen prompt run with the VASP solver configuration. LatticeMind passed every listed case; the direct-LLM comparison is shown in Fig.~\ref{fig:si_vasp}.}
\footnotesize
\setlength{\tabcolsep}{2.5pt}
\renewcommand{\arraystretch}{0.88}
\begin{tabular}{@{}p{0.34\columnwidth}p{0.52\columnwidth}c@{}}
\toprule
Case & Workflow tested & Result \\
\midrule
Ge SCF & SCF (semiconductor) & Pass \\
NaCl SCF & SCF (ionic, 2-species) & Pass \\
Diamond SCF & SCF (wide-gap) & Pass \\
Si DOS & SCF $\rightarrow$ DOS & Pass \\
Fe spin DOS & spin SCF $\rightarrow$ DOS & Pass \\
Diamond band & SCF $\rightarrow$ band & Pass \\
GaAs band & SCF $\rightarrow$ band (III-V) & Pass \\
Al k-mesh & k-mesh scan (3x SCF) & Pass \\
Diamond EOS & EOS scan (3x SCF) & Pass \\
Ta bcc SCF & SCF (bcc metal) & Pass \\
Nb DOS & SCF $\rightarrow$ DOS (bcc metal) & Pass \\
MgB$_2$ DOS & SCF $\rightarrow$ DOS (metal) & Pass \\
Al$_2$O$_3$ DOS & SCF $\rightarrow$ DOS (corundum) & Pass \\
BN band & SCF $\rightarrow$ band (cubic BN) & Pass \\
ZnO band & SCF $\rightarrow$ band (wurtzite) & Pass \\
GaN band & SCF $\rightarrow$ band (wurtzite) & Pass \\
Diamond band(c) & SCF $\rightarrow$ band (compressed) & Pass \\
Diamond strain & SCF $\rightarrow$ band (2 strains) & Pass \\
Fe DOS chain & spin SCF $\rightarrow$ DOS chain & Pass \\
\bottomrule
\end{tabular}
\end{table}

\begin{figure}[!htbp]
\centering
\includegraphics[width=\linewidth]{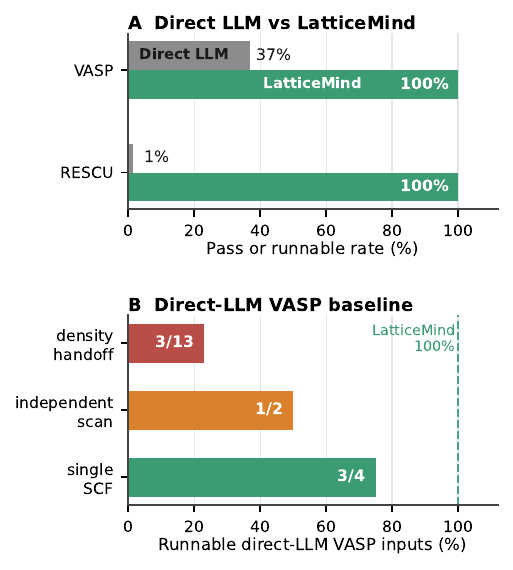}
\caption{\label{fig:si_vasp}
Direct-LLM baseline versus LatticeMind on RESCU and VASP. (a) Gray bars denote one-shot direct-LLM generation with no workflow contract, validation, or repair; green bars denote full LatticeMind with workflow contracts, deterministic validation, and bounded repair. The direct-LLM VASP bar uses the lenient criterion of runnable generated inputs, whereas the LatticeMind bars report validation-gated reference-check success. (b) Direct-LLM VASP results partitioned by workflow family. The few density-handoff inputs that run do so by recomputing each stage rather than reusing the converged density; LatticeMind constructs the required handoff in all 13 density-reuse chains.}
\end{figure}

\subsection{Direct-LLM baseline on VASP}

A natural question is whether the architecture is needed when the target code is widely represented in model training data. We therefore ran a direct-LLM baseline in which the same GPT-5.4-mini backend was asked, in one shot, to produce complete VASP input files with no workflow contract, validation, or repair. Each generated project was scored on whether its inputs ran to completion in VASP, a more lenient bar than the LatticeMind reference checks.

The direct-LLM baseline produced runnable VASP inputs in 7 of 19 cases (37\%), with or without public documentation (Fig.~\ref{fig:si_vasp}). Its successes were self-contained calculations: single-point SCF and independent multi-point scans (the two self-contained failures here were single-run input-generation glitches on the diamond cell, which the model clears in other runs). The multistage density-reuse chains are where it breaks down: across the 13 SCF-to-DOS, SCF-to-band, and spin-polarized chains, the baseline constructed the required cross-stage charge-density handoff in none. The few chains whose inputs ran did so only by recomputing each stage from scratch rather than reusing the converged density; the rest failed outright. Thus stronger raw familiarity with VASP raises the direct-LLM baseline above its RESCU result of at most 1.4\%, but does not by itself produce correct multistage workflows. LatticeMind constructed the handoff in all 13 chains and reached 100\% on the same 19 VASP cases.

\section{Photocatalyst case study}\label{si:casestudy}

\subsection{Setup-reconstruction benchmark}

For the \RhInGaN\ workflow, a compact 12-case file-generation benchmark tested whether LatticeMind could create dependency-consistent RESCU setup files for the main production operations: host and Rh$_6$-cluster SCF setup, slab potential and PDOS handoffs, A/B region and clean-PDOS setup, real-space-grid convergence setup, and replicate-unit setup. Objective checks verified the expected input files, workflow-contract requirements, valid \texttt{rho.in} density handoffs, unique scan output paths, and explicit atom, orbital, layer, or region assignments for projected quantities. The campaign passed all objective checks: 12/12 workflows completed, 23/23 expected RESCU input files generated, and 12/12 cases passed without expert-review flags.

\subsection{Composition-trend descriptors}

Table~\ref{tab:si_summary_metrics} reports the full per-composition photocatalyst descriptor values from the production calculations, summarizing the composition trends discussed in the main text.

\begin{table}[t]
\caption{\label{tab:si_summary_metrics}
Summary of photocatalyst descriptors over independent SQS configurations. Values are means and standard deviations over the configurations where each descriptor is defined; $\DNgap$ and $\DNfermi$ are in states/atom. Gap-window descriptors require a clean Region-B reference gap, so the $x=0.375$ entries use only the one qualifying configuration. $\DNfermi$ integrates the covered-minus-uncovered surface PDOS from $-0.1$ to $+0.1$~eV relative to the Fermi level and is defined for every configuration.}
\small
\begin{tabular}{@{}lccc@{}}
\toprule
Descriptor & $x=0.125$ & $x=0.250$ & $x=0.375$ \\
\midrule
Clean gap & 3/3 & 3/3 & 1/3 \\
$\Egref$ (eV) & $2.84\pm0.18$ & $2.71\pm0.24$ & $2.12$ \\
$\DEC$ (eV) & $0.25\pm0.19$ & $0.49\pm0.04$ & $0.58$ \\
$\DNgap$ & $0.281\pm0.121$ & $0.228\pm0.115$ & $0.339$ \\
$\DNfermi$ & $0.022\pm0.016$ & $0.021\pm0.019$ & $0.039\pm0.044$ \\
$f_d^{\mathrm{Rh}}$ & $0.880\pm0.018$ & $0.889\pm0.005$ & $0.873$ \\
\bottomrule
\end{tabular}
\end{table}

\subsection{Region-B gap-finding protocol}\label{si:gapprotocol}

The Region-B reference gap was located by the same fixed automated rule for every SQS configuration. Each projected curve was smoothed by a reflect-padded Gaussian kernel ($\sigma=2$ energy-grid points). At each candidate interior layer $L=3$--6, the valence and conduction onsets were the highest and lowest energies, respectively, where the N-projected and (Ga+In)-projected PDOS crossed $\alpha\max(\rho)$. The valence and conduction search windows relative to the Fermi level were $[-6,0.5]$ and $[-0.5,6]$~eV. We swept $\alpha\in\{0.005,0.01,0.02,0.03,0.05\}$ and rejected candidate gaps outside $0.5\leq E_g\leq4.5$~eV. The accepted surviving window minimizes
\begin{equation}
\mathrm{score} = I_{\mathrm{gap}}^{\mathrm{frac}} + \frac{0.03}{\max(E_g,\,10^{-3})} - 0.002\,b_{\mathrm{int}}(L),
\end{equation}
Here $I_{\mathrm{gap}}^{\mathrm{frac}}$ is the fraction of the total substrate-projected PDOS integral over $[-10,10]$~eV that lies in the candidate gap; $b_{\mathrm{int}}(L)=1-\min[1,|L-4.5|/1.5]$ for $L=3$--6.

\subsection{Post-processing robustness}\label{si:robustness}

To test how strongly the reported descriptors depend on these post-processing choices, we swept the onset threshold and the interior reference-layer choice and recorded the range of each descriptor over the sweep (Table~\ref{tab:si_robustness}). The covered-minus-uncovered gap-window excess $\DNgap$ and the Rh $d$-fraction $f_d^{\mathrm{Rh}}$ are robust: $\DNgap$ stays positive at every composition across the full sweep (minimum $0.096$~states/atom), and $f_d^{\mathrm{Rh}}$ stays above $0.8$ (minimum $0.857$). The conduction-onset descriptor $\DEC$ is not sign-robust under the sweep, its sign or zero-crossing depending on the onset and reference-layer choice, which is why the main text reports it only as a fixed-protocol descriptor rather than a threshold-independent band-bending measurement. The reference gap $\Egref$ stays positive at every composition, but its absolute value varies with the onset and reference-layer choice, so only the fixed-protocol composition trend is reported.

\begin{table}[t]
\caption{\label{tab:si_robustness}
Robustness of the photocatalyst descriptors to the post-processing onset-threshold and interior reference-layer sweep, across the four sampled compositions. ``Robust'' means the sign (or, for $f_d^{\mathrm{Rh}}$, the $d$-dominance) holds across the full sweep at every composition.}
\footnotesize
\begin{tabular}{@{}l p{0.41\columnwidth} p{0.30\columnwidth}@{}}
\toprule
Descriptor & Behavior across the sweep & Reported as \\
\midrule
$\DNgap$ & Positive at all four (sweep min $0.096$~states/atom) & Robust \\
$f_d^{\mathrm{Rh}}$ & $\geq 0.8$ at all four (sweep min $0.857$) & Robust \\
$\DEC$ & Sign or zero-crossing depends on choices & Fixed-protocol descriptor \\
$\Egref$ & Positive at all four; magnitude varies & Fixed-protocol trend \\
\bottomrule
\end{tabular}
\end{table}

\subsection{Production-level scope}\label{si:prodscope}

These PDOS-derived PBE slab results cover one Rh$_6$ motif, one electronic-structure protocol, and three SQS configurations per alloyed composition. A converged predictive model would require further tests of spin polarization, solvent and temperature effects, reaction barriers, excited states, adsorption motifs, numerical parameters, and alloy configurations; those tests are outside this validation-gated workflow demonstration.